\documentclass[showpacs,superscriptaddress,amsmath,amssymb,nofootinbib]{revtex4-2}
\usepackage{graphicx}
\usepackage{epsfig}
\usepackage{overpic}
\usepackage{dcolumn}
\usepackage{ulem}
\usepackage{bm}
\usepackage{lineno}
\usepackage{xspace}
\usepackage{multirow}
\usepackage{epstopdf}
\usepackage{xcolor}
\usepackage{soul}
\usepackage[colorlinks,allcolors=black]{hyperref}
\usepackage{verbatim}
\usepackage{enumitem}
\usepackage{todonotes}
\usepackage{subcaption}
\usepackage{dcolumn}
\usepackage{bm}
\usepackage{threeparttable}
\usepackage{multirow}
\usepackage[figuresright]{rotating}
\usepackage{makecell}
\usepackage{verbatim}
\usepackage{slashed}
\usepackage{mathtools}
\usepackage{tikz-feynman}
\usepackage{utfsym}
\include{def-com}
\newcommand{\scnt}{\affiliation{Southern Center for Nuclear-Science Theory (SCNT), Institute of Modern Physics, Chinese Academy of Sciences, Huizhou 516000, Guangdong Province, China}}

\newcommand{\half}{\frac12}
\newcommand{\y}{\usym{2714}}
\newcommand{\n}{\usym{2718}}

\begin{document}
\title{\boldmath The Cabibbo-favored hadronic weak decays of the $\Xi_c$ in the quark model}

\author{Peng-Yu Niu}\email{niupy@m.scnu.edu.cn}
\address{Guangdong Provincial Key Laboratory of Nuclear Science, Institute of Quantum Matter, South China Normal University, Guangzhou 510006, China}
\affiliation{Guangdong-Hong Kong Joint Laboratory of Quantum Matter, Southern Nuclear Science Computing Center, South China Normal University, Guangzhou 510006, China}

\author{Qian Wang}\email{qianwang@m.scnu.edu.cn}
\address{Guangdong Provincial Key Laboratory of Nuclear Science, Institute of Quantum Matter, South China Normal University, Guangzhou 510006, China}
\affiliation{Guangdong-Hong Kong Joint Laboratory of Quantum Matter, Southern Nuclear Science Computing Center, South China Normal University, Guangzhou 510006, China}
\scnt

\author{Qiang Zhao}\email{zhaoq@ihep.ac.cn}
\affiliation{Institute of High Energy Physics, Chinese Academy of Sciences, Beijing 100049, China }
\affiliation{University of Chinese Academy of Sciences, Beijing 100049, China}
\affiliation{Center for High Energy Physics, Henan Academy of Sciences, Zhengzhou 450046, China}

\date{\today}

\begin{abstract}
The Cabibbo-favored hadronic weak decay of the $\Xi_c^+\to \Xi^0\pi^+$, $\Xi_c^0\to \Xi^-\pi^+$, $\Xi_c^0\to \Xi^0\pi^0$, $\Xi_c^0\to \Xi^0\eta^{(')}$ and $\Xi_c^0\to \Xi^+K^-$ are studied in the non-relativistic constituent quark model. By analyzing their decay mechanisms at the quark level we show that the pole terms are essential for understanding the transition dynamics in addition to the usually considered direct meson emission process and color suppressed process in the charmed baryon hadronic weak decays. 
The experimentally measurable asymmetry parameters are also predicted in order to further pin down the decay mechanism.
\end{abstract}

\maketitle

\section{Introduction}
The hadronic weak decay of charmed baryons is of great interest since this is the kinematic region where both the strong and weak interactions play a crucial role. On the one hand, the charmed baryon hadronic weak decay can provide a channel for examining the SU(3) flavor symmetry via the combined study of those processes that can be connected by the SU(3) flavor symmetry. On the other hand, one recognizes that the non-factorizable transitions may be non-negligible~\cite{Niu:2020gjw} and more detailed investigations are needed. Different from the hadronic weak decays of bottomed baryons where contributions of the non-factorizable processes can be neglected~\cite{Cheng:2021qpd}, in the charm sector various measurements seem to indicate the non-trivial role played by the non-factorizable processes. Note that more and more data have been accumulated in recent years by different experiments, such as BES III collaboration~\cite{BESIII:2022bkj,BESIII:2022wxj, BESIII:2022izy,BESIII:2022tnm,BESIII:2023ooh,BESIII:2023wrw,BESIII:2024jlj,BESIII:2024cbr}, LHCb collaboration~\cite{LHCb:2022rpd,LHCb:2022mzw,LHCb:2023tma,LHCb:2023ngz,LHCb:2023eeb,LHCb:2024tnq}, Belle and Belle II Collaborations~\cite{Belle-II:2024jql, Belle-II:2024vax}. 
Benefiting from this, a combined study of these transitions should allow us to gain more insights into the mechanism for the hadronic weak decay of charmed baryons.

Recently, the Belle and Belle II Collaborations first reported the absolute branching ratios~\cite{Belle-II:2024jql} of the two-body weak hadronic decays of the $\Xi_c^0$
\begin{align}
\mathrm{Br}(\Xi_c^0 \to \Xi^0 \pi^0)&=(6.9\pm 0.3(\mathrm{stat})\pm 0.5(\mathrm{syst})\pm 1.3(\mathrm{norm})) \times 10^{-3}, \\
\mathrm{Br} (\Xi_c^0 \to \Xi^0 \eta)&=(1.6 \pm 0.2(\mathrm{stat}) \pm 0.2(\mathrm{syst}) \pm 0.3(\mathrm{norm})) \times 10^{-3}, \\
\mathrm{Br} (\Xi_c^0 \to \Xi^0 \eta')&=  (1.2 \pm 0.3(\mathrm{stat}) \pm 0.1(\mathrm{syst}) \pm 0.2(\mathrm{norm})) \times 10^{-3} ,
\end{align}
where the first and second uncertainties are the statistic and systematic ones, respectively. The third errors are from the uncertainty of the branching ratio of $ \Xi_c^0 \to \Xi^- \pi^+$\cite{ParticleDataGroup:2022pth} which is  
\begin{align}
\mathrm{\Xi_c^0 \to \Xi^- \pi^+}=( 1.43\pm 0.27)\% ,
\end{align}
where serves as a normalization quantity.
The asymmetry parameter of $\Xi_c^0 \to \Xi^0 \pi^0 $ was also measured by Belle-II~\cite{Belle-II:2024jql}
\begin{align}
\alpha(\Xi_c^0 \to \Xi^0 \pi^0)&= -0.90 \pm 0.15(\mathrm{stat}) \pm 0.23(\mathrm{syst}),
\end{align}
which is important for shedding light on the underlying dynamics.  
Although there are several theoretical calculations of these branching ratios, 
none of them is in full agreement with the experimental data as presented in Ref.~\cite{Belle-II:2024jql}. 
 
Motivated by the above observations, 
we focus on the Cabibbo-favored two-body hadronic weak decay $\Xi_c\to \mathcal B \mathcal P$,  
where $\mathcal B$ stands for the $J^p=\frac{1}{2}^+$ strange baryons and $\mathcal P$ stands for pseudoscalar mesons. For the Cabibbo-favored two-body hadronic weak decay of the $\Xi_c$, the $s$ quark in the $\Xi_c$ should not be involved in the weak interaction. Thus, some decay diagrams are forbidden which provides an golden opportunity to study the decay mechanisms, especially, to clarify the contributions of the non-factorizable diagrams.

As mentioned at the beginning, more and more charmed baryons decay modes, branching ratios and the other physical observables have been measured or updated.
Taking advantage of these experimental data one can perform a data-based study. 
For instance, global-fit in tandem with the irreducible flavor symmetry approach or topological-diagram approach for the charmed baryon nonleptonic weak decays can be found in the literature~\cite{Sharma:1996sc,Geng:2018plk,Geng:2019xbo, Jia:2019zxi,Zhao:2018mov,Hsiao:2021nsc,Huang:2021aqu,Xing:2023dni,Zhong:2022exp, Geng:2023pkr,Liu:2023dvg,Wang:2024ztg,Xing:2024nvg,Zhong:2024qqs, Zhong:2024zme}.
This method embeds the clear physical picture and is also a powerful method for predicting new decay modes. 
However, none of them can obtain overall consistent results for all of the 
branching ratios. This suggests that some physical dynamics may be missed in the above mentioned method and more elaborate treatment should be taken into account in the charm sector. Dynamic model calculations include various quark models~\cite{Korner:1992wi,Cheng:1993gf,Ivanov:1997ra,Liu:2023pyk}, pole model~\cite{Xu:1992vc, Zenczykowski:1993jm, Zou:2019kzq}, and current algebra approach~\cite{Sharma:1998rd}. Here, only the references that refer to the hadronic weak decay of the $\Xi_c$ are listed.

In this work, the dynamic mechanisms for the two-body hadronic weak decays are presented in the non-relativistic constituent quark model~\cite{Close:1979bt, LeYaouanc:1988fx}. As shown in Refs.~\cite{Niu:2020gjw,Niu:2021qcc}, apart from the direct meson emission (DME) process and the color suppressed (CS) process, where only the short-ranged weak interactions are involved, the pole term transitions, which involve both weak and strong interactions non-locally, turn out to be important in the charm sector. The DME process is factorizable, while the CS process and pole term process are nonfactorizable. These nonfactorizable processes should be dealt with carefully. For instance, it has been shown that the pole terms in the heavy-quark-conserving hadronic weak decay of $\Xi_c$ is the key for understanding the strongly enhanced decay channel $\Xi_c^0\to \Lambda_c\pi^-$~\cite{Niu:2021qcc}.

Our paper is organized as follows. The framework is presented in Sec.~\ref{sec:framework}. The third section is devoted to the results and discussions. 
A brief summary is given in the last section. The wave functions of the relevant hadrons and some necessary materials are listed in the appendix.

\section{Framework}
\label{sec:framework}
We carry out the study in the framework of non-relativistic constituent quark model~\cite{Copley:1979wj,Godfrey:1985xj, Capstick:1986ter}  which has been adopted  successfully for the hadronic weak decays of $\Lambda_c$, $\Xi_c^{+/0}$ and $\Xi_b^{-/0}$~\cite{Niu:2020gjw, Niu:2021qcc}. Here, we only briefly introduce the framework. One advantage of this approach is that one can consider the non-factorizable processes consistently at the quark level, and their interfering effects can be properly included.

\subsection{Operators in the quark level}
We categorize the hadronic weak decay processes into two classes. The first one includes either the $W$-emission or $W$-exchange diagram and only the weak interaction operators are considered explicitly at leading order. The second one involves both the weak and strong interactions explicitly. In such a transition the weak and strong interactions occur non-locally, and pole structures in the transition amplitudes can be identified. Intermediate baryon states may have non-negligible contributions given that the propagating energy is not far away from their masses~\cite{Niu:2021qcc}.
Concerning the production of light pseudoscalar and vector mesons in the charmed baryon hadronic weak decays, the strong interaction between the light meson and constituent quarks can be described in the chiral quark model~\cite{Manohar:1983md, Zhao:2002id, Zhong:2007gp}. 

\subsubsection{The non-relativistic form of the weak interaction}
\begin{figure}[htbp!]
\begin{center}
\begin{subfigure}[htbp!]{0.45\textwidth}
\begin{tikzpicture}[line width=0.6pt]
\begin{feynman}
\vertex (a1) {$ d $};
\vertex[right=2cm of a1] (a2);
\vertex[right=2cm of a2] (a3){$ u $};
\vertex[below=4em of a1] (b1) {$ c $};
\vertex[below=4em of a2] (b2) ;
\vertex[below=4em of a3] (b3) {$ s $};
\diagram* {
(a1) -- [fermion] (a2) -- [fermion] (a3),
(b1) -- [fermion] (b2) -- [fermion] (b3),
(a2) -- [boson, edge label=$W$] (b2),
};
\end{feynman}
\end{tikzpicture}
\caption{}
\end{subfigure}
~
\begin{subfigure}[htbp]{0.45\textwidth}
\begin{tikzpicture}[line width=0.6pt]
\begin{feynman}
\vertex (a1) {$ c $};
\vertex[right=1.5cm of a1] (a2);
\vertex[right=1cm of a2] (a3);
\vertex[right=1.5cm of a3] (a4){$ s $};
\vertex[above=2em of  a3] (b1);
\vertex[above=2.5em of a4] (b2) {$u$};
\vertex[above=4em of a4] (b3) {$ \bar d $};
\diagram* {
(a1) -- [fermion] (a2),
(a2) -- [fermion] (a4),
(b1) -- [fermion] (b2),
(b3) -- [fermion] (b1),
(a2) -- [boson, edge label=$W$] (b1),
};
\end{feynman}
\end{tikzpicture}
\caption{}
\end{subfigure}
\caption{The transition of the $cd\to su$ (left panel) and $c \to u \bar d s$ (right panel) processes.}
\label{fig:cdus}
\end{center}
\end{figure}
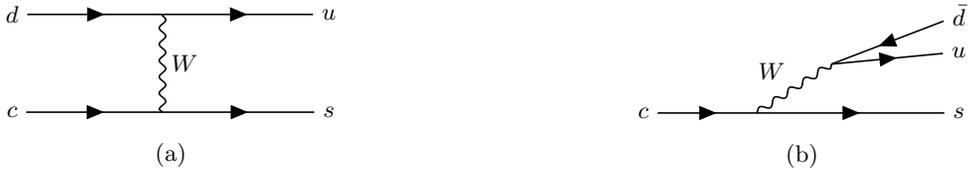

In the limit of low momentum transfer, the weak interaction can be described by the four-fermion interaction~\cite{LeYaouanc:1978ef, LeYaouanc:1988fx} with the weak couplings and gauge boson mass encoded in the Fermi constant $G_F$. 
In this sense, the effective Hamiltonian can be expressed as:
\begin{align}
H_W=\frac{G_F}{\sqrt 2}\int d \bm x \frac12 \{ J^{-,\mu}(\bm x),J^{+}_{\mu}(\bm x) \}.
\end{align}
 $J^{\pm}_\mu$ is the quark current and can be written as 
\begin{align}
J^{+,\mu}(\bm x)&=
\begin{pmatrix}\bar u&\bar c \end{pmatrix}
\gamma^\mu(1-\gamma_5)
\begin{pmatrix}\cos \theta_C & \sin \theta_C \\ -\sin \theta_C &\cos \theta_C \end{pmatrix} \begin{pmatrix} d\\s \end{pmatrix}, \\
J^{-,\mu}(\bm x)&=
\begin{pmatrix}\bar d &\bar s \end{pmatrix} \begin{pmatrix}\cos \theta_C & -\sin \theta_C \\ \sin \theta_C &\cos \theta_C \end{pmatrix} \gamma^\mu(1-\gamma_5)
\begin{pmatrix} u\\c \end{pmatrix},
\end{align}
where $\theta_C$ is the Cabibbo angle. With this Hamiltonian, one can extract the operators for the $cd\to s u$ and $c \to u \bar d s$ transitions, as illustrated by Fig.~\ref{fig:cdus}. 
The corresponding Cabibbo-favored  hadronic weak decay of the $\Xi_c$ in quark level can be achieved as shown by the Feynman diagram presented in App.~\ref{app:figamp}. In this case, the weak transition operators are written as  
\begin{align}
H_{W,cd\to su}&=\frac{G_F}{\sqrt{2}} V_{u d} V_{c s} \frac{1}{(2 \pi)^3} \delta^3\left(\boldsymbol{p}_c+\boldsymbol{p}_d-\boldsymbol{p}_u-\boldsymbol{p}_s\right) \bar{u}\left(\boldsymbol{p}_s\right) \gamma_\mu\left(1-\gamma_5\right) u\left(\boldsymbol{p}_c\right) \bar{u}\left(\boldsymbol{p}_d\right) \gamma^\mu\left(1-\gamma_5\right) u\left(\boldsymbol{p}_u\right)\hat{\alpha}^{(-)} \hat{\beta}^{(+)}, \\
H_{W,c\to u \bar d s}&=\frac{G_F}{\sqrt{2}} V_{u d} V_{c s} \frac{1}{(2 \pi)^3} \delta^3\left(\boldsymbol{p}_c-\boldsymbol{p}_d-\boldsymbol{p}_u-\boldsymbol{p}_s\right) \bar{u}\left(\boldsymbol{p}_s\right) \gamma_\mu\left(1-\gamma_5\right) u\left(\boldsymbol{p}_c\right) \bar{u}\left(\boldsymbol{p}_u\right) \gamma^\mu\left(1-\gamma_5\right) v\left(\boldsymbol{p}_{\bar d}\right) \hat{\alpha}^{(-)} \hat{I}^{\prime}_p.
\end{align}
where $V_{cs}$ and $V_{ud}$ are the Cabibbo-Kobayashi-Maskawa (CKM) matrix elements. $\hat\alpha$ and $\hat\beta$ are the flavor-changing operators with $\hat\alpha^{(-)}c=s,~\hat\beta^{(+)}d=u$. $\hat I'_p$ is the isospin operator for the pseudoscalar meson production for the color suppressed process. 
Its explicit form is 
\begin{align}
\hat I'_p=\begin{cases}
b^\dagger_u b_u                    &\mathrm{for}~ \pi^+,\\
-\frac{1}{\sqrt 2}b^\dagger_u b_d  &\mathrm{for}~ \pi^0,\\
\frac{1}{\sqrt 2}\left[ b^\dagger_u b_d+ b^\dagger_d b_u\right]   &\mathrm{for}~ \eta_q.
\end{cases}
\end{align}
$b^\dagger_{u/d}$ and $b_{u/d}$ are the creation and annihilation operators of the $u/d$ quark.

According to the number of $\gamma_5$, $H_W$ can always be separated into two parts, i.e. the parity-conserving (PC) one with even number of $\gamma_5$ and parity-violating (PV) one with odd number of $\gamma_5$, labeled as $H^{\mathrm{PC}}_{W}$ and $H^{\mathrm{PV}}_{W}$, respectively. In the non-relativistic limit, for the $cd\to su$ process, the transition operator $H_{W,cd \to su }$ is rewritten as:
\begin{align}
\label{eq:HW22}
H_{W,cd\to s u}^{\mathrm{P C}}&= \frac{G_F}{\sqrt{2}} V_{u d} V_{c s} \frac{1}{(2 \pi)^3} \sum_{i \neq j} \hat{\alpha}_i^{(-)} \hat{\beta}_j^{(+)} \delta^3\left(\boldsymbol{p}_i^{\prime}+\boldsymbol{p}_j^{\prime}-\boldsymbol{p}_i-\boldsymbol{p}_j\right)\left(1-\left\langle s_{z, i}^{\prime}\left|\boldsymbol{\sigma}_{\boldsymbol{i}}\right| s_{z, i}\right\rangle\left\langle s_{z, j}^{\prime}\left|\boldsymbol{\sigma}_j\right| s_{z, j}\right\rangle\right), \\
H_{W, cd\to s u}^{\mathrm{P V}}&= \frac{G_F}{\sqrt{2}} V_{u d} V_{c s} \frac{1}{(2 \pi)^3} \sum_{i \neq j} \hat{\alpha}_i^{(-)} \hat{\beta}_j^{(+)} \delta^3\left(\boldsymbol{p}_i^{\prime}+\boldsymbol{p}_j^{\prime}-\boldsymbol{p}_i-\boldsymbol{p}_j\right)  \notag  \\
&\times\left\{-\left(\left\langle s_{z, i}^{\prime}\left|\boldsymbol{\sigma}_{\boldsymbol{i}}\right| s_{z, i}\right\rangle-\left\langle s_{z, i}^{\prime}\left|\boldsymbol{\sigma}_j\right| s_{z, j}\right\rangle\right)\left[\left(\frac{\boldsymbol{p}_i}{2 m_i}-\frac{\boldsymbol{p}_j}{2 m_j}\right)+\left(\frac{\boldsymbol{p}_i^{\prime}}{2 m_i^{\prime}}-\frac{\boldsymbol{p}_j^{\prime}}{2 m_j^{\prime}}\right)\right]\right. \notag \\
& \left.+i\left(\left\langle s_{z, i}^{\prime}\left|\boldsymbol{\sigma}_i\right| s_{z, i}\right\rangle \times\left\langle s_{z, i}^{\prime}\left|\boldsymbol{\sigma}_j\right| s_{z, j}\right\rangle\right)\left[\left(\frac{\boldsymbol{p}_i}{2 m_i}-\frac{\boldsymbol{p}_j}{2 m_j}\right)-\left(\frac{\boldsymbol{p}_i^{\prime}}{2 m_i^{\prime}}-\frac{\boldsymbol{p}_j^{\prime}}{2 m_j^{\prime}}\right)\right]\right\},
\end{align}
The subscripts $i$ and $j$ are used to indicate the operators acting on the $i$th and $j$th quarks. Similarly, with the quark labels shown in Fig.~\ref{fig:cs}, the operators $H_{c \to u \bar d s }$ are written as:
\begin{align}
\label{eq:HW13}
H_{W,c \to u\bar d s}^{\mathrm{P C}}&=\frac{G_F}{\sqrt{2}} V_{u d} V_{c s} \frac{\beta}{(2 \pi)^3} \delta^3\left(\boldsymbol{p}_3-\boldsymbol{p}_3^{\prime}-\boldsymbol{p}_4-\boldsymbol{p}_5\right)\left\{\left\langle s_3^{\prime}|I| s_3\right\rangle\left\langle s_5 \bar{s}_4|\boldsymbol{\sigma}| 0\right\rangle\left(\frac{\boldsymbol{p}_5}{2 m_5}+\frac{\boldsymbol{p}_4}{2 m_4}\right)\right.  \notag  \\
& -\left[\left(\frac{\boldsymbol{p}_3^{\prime}}{2 m_3{ }^{\prime}}+\frac{\boldsymbol{p}_3}{2 m_3}\right)\left\langle s_3^{\prime}|I| s_3\right\rangle-i\left\langle s_3^{\prime}|\boldsymbol{\sigma}| s_3\right\rangle \times\left(\frac{\boldsymbol{p}_3}{2 m_3}-\frac{\boldsymbol{p}_3^{\prime}}{2 m_3^{\prime}}\right)\right]\left\langle s_5 \bar{s}_4|\boldsymbol{\sigma}| 0\right\rangle  \notag  \\
& -\left\langle s_3^{\prime}|\boldsymbol{\sigma}| s_3\right\rangle\left[\left(\frac{\boldsymbol{p}_5}{2 m_5}+\frac{\boldsymbol{p}_4}{2 m_4}\right)\left\langle s_5 \bar{s}_4|I| 0\right\rangle-i\left\langle s_5 \bar{s}_4|\boldsymbol{\sigma}| 0\right\rangle \times\left(\frac{\boldsymbol{p}_4}{2 m_4}-\frac{\boldsymbol{p}_5}{2 m_5}\right)\right]   \notag \\
& \left.+\left\langle s_3^{\prime}|\boldsymbol{\sigma}| s_3\right\rangle\left(\frac{\boldsymbol{p}_3^{\prime}}{2 m_3^{\prime}}+\frac{\boldsymbol{p}_3}{2 m_3}\right)\left\langle s_5 \bar{s}_4|I| 0\right\rangle\right\} \hat{\alpha}_3^{(-)} \hat{I}_p^{\prime},    \\
H_{W,c \to u\bar d s}^{\mathrm{P V}}&= \frac{G_F}{\sqrt{2}} V_{u d} V_{c s} \frac{\beta}{(2 \pi)^3} \delta^3\left(\boldsymbol{p}_3-\boldsymbol{p}_3^{\prime}-\boldsymbol{p}_4-\boldsymbol{p}_5\right)\left(-\left\langle s_3^{\prime}|I| s_3\right\rangle\left\langle s_5 \bar{s}_4|I| 0\right\rangle+\left\langle s_3^{\prime}|\boldsymbol{\sigma}| s_3\right\rangle\left\langle s_5 \bar{s}_4|\boldsymbol{\sigma}| 0\right\rangle\right) \hat{\alpha}_3^{(-)} \hat{I}_p^{\prime}.
\end{align}
Here $I$ is the two-dimension unit matrix. Note that we have specify the quark which the operator acts on. Thus, there is a symmetry factor $\beta$ which equals to $1$ for the DME process and $2/3$ for the CS process, which stems from the anti-symmetric baryon wave functions.
$\bar s_4$ is the spin of the $4$th particle, e.g. an anti-quark. In order to evaluate the spin matrix element containing an anti-quark conveniently, the particle-hole conjugation~\cite{Racah:1942gsc} is employed here:
\begin{align}
\langle j,-m| \to(-1)^{j+m}|j,m\rangle.
\end{align} 

\subsubsection{The non-relativistic form of the quark-meson interaction}
At leading order, the effective Hamiltonian in chiral quark model~\cite{Manohar:1983md, Zhong:2007gp} is written as:
\begin{align}
\label{equ:Hpi}
H_{\chi}=\int d \bm x \frac{1}{f_m}\bar q(\bm x) \gamma_\mu \gamma_5 q(\bm x) \partial^\mu \phi_m(\bm x) ,
\end{align}
where $f_m$ is the pseudoscalar meson decay constant and the relation between these constants can be found in Re.~\cite{Feldmann:1998vh,Feldmann:1998sh}. $q(\bm x)$ is the light quark field and $\phi_m$ represents the nonet pseudoscalar meson field, which is expressed as 
\begin{align}
\phi_m=\left(\begin{array}{ccc}
\frac{1}{\sqrt 2}(\pi^0+\cos\phi \eta+\sin\phi \eta') & \pi^+ & K^+ \\
\pi^- & -\frac{1}{\sqrt 2}(-\pi^0+\cos\phi \eta+\sin\phi \eta')  & K^0\\
K^-   &\bar K^0  & -\sin\phi \eta+\cos\phi \eta'
\end{array}\right).
\end{align}
where the mixing angle between $\eta$ and $\eta^\prime$ is $\phi=39.3^\circ$\cite{ParticleDataGroup:2022pth}.
The non-relativistic form of $H_m$ used in the calculation is expressed as:
\begin{align}
H_{\chi}=\frac{1}{\sqrt{(2\pi)^3 2\omega_m}} \sum_{j=1}^3 \frac{1}{f_m}\left[\omega_m\left(\frac{\bm\sigma\cdot \bm p^j_f}{2m_f}+\frac{\bm\sigma\cdot \bm p^j_i}{2m_i}\right) -\bm \sigma \cdot \bm k\right] \hat I^j_\chi \delta^3(\bm p^j_f+\bm k-\bm p_i^j),
\end{align}
where $\omega_m$ and $\bm k$ are the energy and momentum of the pseudoscalar meson, respectively. 
$\bm p^j_i$ and $\bm p^j_f$ are the initial and final momentum of $j$th quark, respectively. $\hat I^j_\chi$ is the corresponding isospin operator acting on the $j$th quark and can be written as:
\begin{align}
\hat I^j_\chi=\begin{cases}
b^\dagger_d b_u,                      &\mathrm{for}~ \pi^+,\\
\frac{1}{\sqrt 2}\left [b^\dagger _u b_u - b^\dagger_d b_d \right],  &\mathrm{for}~ \pi^0,\\
b^\dagger_s b_u,                      &\mathrm{for}~ K^+,\\
b^\dagger_u b_s,                      &\mathrm{for}~ K^-, \\
\frac{\cos\phi}{\sqrt 2}\left[ b^\dagger_u b_u+b^\dagger_d b_d\right]- \sin \phi b^\dagger_s b_s ,                     &\mathrm{for}~ \eta, \\
\frac{\sin\phi}{\sqrt 2}\left[ b^\dagger_u b_u+b^\dagger_d b_d\right]+\cos \phi b^\dagger_s b_s ,                     &\mathrm{for}~ \eta'.
\end{cases}
\end{align}

\subsection{Amplitudes, asymmetry parameter and decay width}
All the decay processes are calculated in the rest frame of initial charmed baryons. We set the momentum direction of the final baryons as the $z$ axis. The amplitudes of the DME, CS and Pole terms are constructed as follows:
\begin{align}
\mathcal M_{\mathrm{CS},\mathrm{PC/PV}}^{J_f,J_f^z;J_i,J_i^z}&=\langle B_f(\bm P_f;J_f,J_f^z);M(\bm k)|H^{\mathrm{PC/PV}}_{W,c\to u \bar b s}|B_c(\bm P_i;J_i,J_i^z) \rangle,\\
\mathcal M_{\mathrm{DME},\mathrm{PC/PV}}^{J_f,J_f^z;J_i,J_i^z}&=\langle B_f(\bm P_f;J_f,J_f^z);M(\bm k)|H^\mathrm{PC/PV}_{W,c\to u \bar b s}|B_c(\bm P_i;J_i,J_i^z) \rangle, \\
\mathcal M_\mathrm{Pole}^{J_f,J_f^z;J_i,J_i^z}&=\mathcal M_{\mathrm{Pole};\mathrm{PC}}^{J_f,J_f^z;J_i,J_i^z}+
\mathcal M_{\mathrm{Pole};\mathrm{PV}}^{J_f,J_f^z;J_i,J_i^z}.
\end{align}
$B_i(\bm P_i;J_i,J^z_i)$ and $B_f(\bm P_f;J_f,J^z_f)$ are the initial baryon and final baryon states, respectively, with their four momentum, spin and the third component of the spin presented explicitly. The pseudoscalar meson is labeled as $M(\bm k)$.
The amplitude of the CS process has the similar form of that of the DME process, but with the different momentum conservation condition which is listed in the App.~\ref{app:cal}. This is due to the different constituent quarks of the final meson which are involved in the transition.
For the pole terms, there are two kinds of processes as mentioned early and the corresponding amplitudes read as:
\begin{align}
&\mathcal M_{\mathrm{WS};\mathrm{PC/PV}}^{J_f,J_f^z;J_i,J_i^z}\notag  \\
&=\langle B_f(\bm P_f;J_f,J_f^z);M(\bm k) | H_\chi | B_m(\bm P_i;J_i,J_i^z) \rangle \frac{i}{ \slashed p_{B_m}-m_{B_m} +i \frac{\Gamma_{B_m}}{2} } \langle B_m(\bm P_i;J_i,J_i^z) | H^{\mathrm{PC/PV}}_{W,cd \to su} | B_i(\bm P_i;J_i,J_i^z) \rangle,
\\
&\mathcal M_{\mathrm{SW};\mathrm{PC/PV}}^{J_f,J_f^z;J_i,J_i^z}\notag  \\
&= \langle B_f(\bm P_f;J_f,J_f^z) | H^{\mathrm{PC/PV}}_{W,cd \to su} | B'_c(\bm P_i;J_i,J_i^z) \rangle  \frac{i}{ \slashed p_{B'_c}-m_{B'_c} +i \frac{\Gamma_{B'_c}}{2} } \langle B'_c(\bm P_i;J_i,J_i^z);M(\bm k) | H_\chi | B_c(\bm P_i;J_i,J_i^z) \rangle.
\end{align}
where $|B_m(\bm P_i;J_i,J_i^z)\rangle$ and $ |B'_c(\bm P_i;J_i,J_i^z)\rangle$ denote the intermediate light baryons and charmed baryons, respectively. All the transition amplitudes contain the information of the flavor, spin and spatial space.

With the normalization convention 
\begin{align}
\langle B_f(P_f) |B_i(P_i) \rangle =\delta^3(\bm P_f- \bm P_i),
\end{align}
the decay width is expressed as
\begin{align}
\Gamma(A\to B+C)=8\pi^2\frac{|\bm k|E_B E_C}{M_A}\frac{1}{2 J_A+1} \sum_\mathrm{spin} \left(|\mathcal M_{\mathrm{PC}}|^2+ |\mathcal M_{\mathrm{PV}}|^2 \right),
\end{align}
where $J_A$ is the spin of the initial state.

Generally, the amplitude for a  $\frac{1}{2}^+$ charmed baryon decaying into a $\frac{1}{2}^+$ strange baryon and a spinless meson can be written in the form~\cite{ParticleDataGroup:2022pth}
\begin{align}
M=G_F m_\pi^2 \bar B_f(A-B\gamma_5) B_i,
\end{align}
where $A$ and $B$ are constants~\cite{Commins:1983ns} in front of the unit matrix and $\gamma_5$ Dirac matrix. The parity conserving and violating amplitudes are proportional to $A$ and $B$, respectively.
The parity asymmetry parameter 
\begin{align}
    \alpha= \frac{2\mathrm{Re}\left[s^* p\right]}{|s|^2+|p|^2},
\end{align}
is one of the most important parameters of baryon decay ~\cite{ParticleDataGroup:2022pth}, which was defined by Lee and Yang~\cite{Lee:1957qs} in the study of parity violation~\cite{LHCb:2024tnq}.
Here
\begin{align}
s=A,\quad p=B\frac{|\bm p_f|}{E_f+m_f}.
\end{align}
With these definitions, one can find re-express the parity asymmetry parameter as:
\begin{align}
\label{eq:ap}
\alpha=\frac{2 \operatorname{Re}\left[\mathcal M^*_\mathrm{P V} \mathcal M_\mathrm{P C}\right]}{\left|\mathcal M_\mathrm{P C}\right|^2 +\left|\mathcal M_\mathrm{P V}\right|^2 } \ ,
\end{align}
where $\mathcal M_{P c}$ and $\mathcal M_{P V}$ are the PC and the PV amplitudes, respectively, with $s_i^z=s_i^z=1/2$ in the quark model.

\section{Results and discussion}

\subsection{Parameters and inputs}

For the quark model calculations, there is a tricky issue that the parameters in the Hamiltonian, for instance the quark masses, as shown in Tab.~\ref{tab:quarkMass}, are different in various works, leaving the final results parameter-dependent. In this work, a typical set of quark masses is used (see the last line of Tab.~\ref{tab:quarkMass}) and its deduced uncertainties will be discussed afterwards.
The spring constants $K$ which describe the two-quark potential are listed in Tab.~\ref{tab:K}. Here different values of $K$ depend on the quark flavor, i.e. the strangeness and charmness of baryons, obtaining the typical value of harmonic oscillator strengths. The uncertainties caused by these parameters will be discussed later.

The mass, decay constants, and harmonic oscillator strengths $R$ for light mesons are listed in Tab.~\ref{tab:meson}. The wave functions of these mesons are listed in App.~\ref{app:wavefunction}. The decay constant $f_q \ (f_s)$ is taken into account if $\eta^{\prime}$ is produced from the $q\bar q(s\bar s)$ quark pair.
The masses of the light baryons and charmed baryons used in our calculation are listed in Tab.~\ref{tab:Lmass} and Tab.~\ref{tab:Cmass}, respectively. The masses of light baryons and the ground states of charmed baryons are taken from PDG~\cite{ParticleDataGroup:2022pth}. However, most of the first excited states of single charmed baryons are not confirmed by experiment. Thus, the predicted values from various quark models~\cite{Yoshida:2015tia,Chen:2016iyi,Roberts:2007ni,Ebert:2011kk,Shah:2016mig,Garcia-Tecocoatzi:2022zrf,Ortiz-Pacheco:2023kjn,Bijker:2020tns} are used to estimate the masses of the first excited states of the charmed baryons. Note that, for a given state, the difference among various models can be significant. As an estimation, we set $20\%$ uncertainties to all the quark model parameters except for the masses of the intermediate spin-parity $1/2^-$ heavy baryons. For a comparison, the masses of these states in the selected works are listed in Tab.~\ref{tab:Cmass}, with the last one adopted in our work. 

\begin{table}[htbp]
\centering
\caption{The quark masses used in different works. The values, both central values and the uncertainties ($20\%$ of the center values), in the last line are used in this work.}
\begin{ruledtabular}
\begin{tabular}{cccc}
Inputs                     &$m_q$ (GeV) &$m_s$ (GeV) &$m_c$ (GeV) 
\\
\hline
Ref.~\cite{Yoshida:2015tia}&$0.30$        &$0.51$        &$1.75$        
\\
Ref.~\cite{Chen:2016iyi}\footnotemark[1]  &$(0.225,0.33)$&$$(0.405,0.45)$$  &$1.68$    
\\
Ref.~\cite{Roberts:2007ni} &$0.28$        &$0.56$        &$1.82$        
\\
Ref.~\cite{Ebert:2011kk}   &$0.33$        &$0.50$        &$1.55$        
\\
Ref.~\cite{Shah:2016mig}   &$0.338$\footnotemark[2]       &$0.50$        &$1.275$       
\\
Ref.~\cite{Garcia-Tecocoatzi:2022zrf} &$0.284$       &$0.455$        &$1.606$       
\\
Ref.~\cite{Ortiz-Pacheco:2023kjn} &$0.292$       &$0.461$        &$1.607$       
\\
Used                       &$0.30\pm0.06$ &$0.50\pm0.10$ &$1.80\pm0.36$ 
\footnotetext[1]{~The mass of light quark is obtained from the mass of light quark cluster $m_{[qq]}$ and $m_{\{qq\}}$.}
\footnotetext[2]{~The mass of $m_u$.}
\end{tabular}\end{ruledtabular}
\label{tab:quarkMass}
\end{table}

\begin{table}[htbp]
\centering
\caption{The spring constant $K$ and the harmonic oscillator strengths $\alpha_\rho$ and $\alpha_\lambda$. The definition of $\alpha_\rho$ and $\alpha_\lambda$ can be find in App.~\ref{app:wavefunction}. $S$ and $C$ are used to label the strangeness and charmness of the baryons, respectively. The uncertainties of $\alpha_\rho$ and $\alpha_\lambda$ are from the $20\%$ uncertainties of  quark masses and $K$. }
\begin{ruledtabular}
\begin{tabular}{ccccc}
system                 &$S=1,C=0$ &$S=2,C=0$  &$S=0,C=1$  &$S=1,C=1$  \\
\hline
$K(\mathrm{GeV}^3)$    &$0.06\pm0.012$ &$0.03\pm0.006$ &$0.02\pm0.004$  &$0.02\pm0.004$ \\
$\alpha_\rho$ (GeV)    &$0.48\pm0.034$ &$0.43\pm0.027$ &$0.37\pm0.04$   &$0.39\pm0.04$  \\
$\alpha_\lambda$ (GeV) &$0.52\pm0.032$ &$0.45\pm0.030$ &$0.44\pm0.05$   &$0.47\pm0.05$  \\
\end{tabular}\end{ruledtabular}
\label{tab:K}
\end{table}

\begin{table}[htbp]
\centering
\caption{The masses, decay constants and $R$s that are the parameters of spatial wave function for pseudoscalar mesons.}
\begin{ruledtabular}
\begin{tabular}{cccccc}
meson               &$\pi^\pm$  &$\pi^0$   &$\eta$    &$\eta'$     &$K^\pm$   \\
\hline
Mass\cite{ParticleDataGroup:2022pth} (MeV)           
                     &$139.57$ &$134.98$  &$547.86$   &$957.78$    &$493.68$  \\
Decay constant\cite{Gan:2020aco} (MeV) &$92.4$   &$92.4$    &$f_q=1.07f_\pi$        &$f_s=1.34f_\pi$   &$1.2 f_\pi$  \\
R (GeV)              &$0.28$   &$0.28$    &$0.4$       &$0.9$         &$0.5$ 
\end{tabular}\end{ruledtabular}
\label{tab:meson}
\end{table}

\begin{table}[htbp!]
\caption{The central values of the light baryons mass and width taken from PDG~\cite{ParticleDataGroup:2022pth} (in unit of GeV). The quantum number of $\Xi(1620)$ and $\Xi(1690)$ are not confirmed and we treat them as the $[70,^2 8]$ and $[70,^4 8]$ states in the SU(6) quark model, respectively\cite{Niu:2020aoz}. }
\begin{ruledtabular}
\begin{tabular}{lcccccc}
Particles  &  $\Xi^0$ & $\Xi^0(1620)$ & $\Xi^0(1690)$ & $\Xi^-$ &   $\Sigma^+$ &  $\Omega^-$   \tabularnewline
\hline
$I(J^P)$ &$\frac{1}{2}^+$ &$\frac{1}{2}^-$&$\frac{1}{2}^-$ &$\frac{1}{2}^+$  & $\frac{1}{2}^+$ & $\frac{3}{2}^+$  \tabularnewline
mass  &1.315 &$\approx 1.620$ &$\approx 1.690$       &1.322 &1.189 &1.672\tabularnewline
width & -    & 0.032          &0.020\footnotemark[1] & -    & -    & - 
\footnotetext[1]{~The value of mixed charges.}
\end{tabular}\end{ruledtabular}
\label{tab:Lmass}
\end{table}

\begin{table}[htbp]
\centering
\caption{The masses of the charmed baryons (in unit of GeV). Only their central values are listed.}
\begin{ruledtabular}
\begin{tabular}{l|l|llll|llll|llll|llll}
\multicolumn{1}{c|}{\multirow{2}[0]{*}{States}}&$\Xi^+_c$ & \multicolumn{4}{l|}{$\Xi^0_c$} & \multicolumn{4}{l|}{$\Xi'^0_c$} & \multicolumn{4}{l|}{$\Sigma_c^+$} & \multicolumn{4}{l}{$\Lambda_c^+$} \\ 
&$|^2 S\rangle$
&$|^2 S\rangle$ &$|^2 P_\lambda\rangle$ &$|^2 P_\rho\rangle$ &$|^4 P_\rho\rangle$
&$|^2 S\rangle$ &$|^2 P_\lambda\rangle$ &$|^2 P_\rho\rangle$ &$|^4 P_\lambda\rangle$ 
&$|^2 S\rangle$ &$|^2 P_\lambda\rangle$ &$|^2 P_\rho\rangle$ &$|^4 P_\lambda\rangle$ 
&$|^2 S\rangle$ &$|^2 P_\lambda\rangle$ &$|^2 P_\rho\rangle$ &$|^4 P_\rho\rangle$ \\
\hline
PDG~\cite{ParticleDataGroup:2022pth} 
&2.468
& 2.470 &$\cdots$&$\cdots$&$\cdots$  
& 2.579 &$\cdots$&$\cdots$&$\cdots$
& 2.453 &$\cdots$&$\cdots$&$\cdots$
& 2.286 &$\cdots$&$\cdots$&$\cdots$ \\
Ref.~\cite{Yoshida:2015tia} 
&$\cdots$   
&$\cdots$&$\cdots$&$\cdots$&$\cdots$ 
&$\cdots$&$\cdots$&$\cdots$&$\cdots$  
&2.460 & 2.802  & 2.909 & 2.826
&2.285 & 2.628 & 2.890  & 2.933\\
Ref.~\cite{Chen:2016iyi}
& 2.470
& 2.470 & 2.793 &$\cdots$& $\cdots$     
& 2.579 & 2.839 &$\cdots$& 2.900   
&2.456&2.795&$\cdots$&2.805       
&2.286&2.614&$\cdots$&$\cdots$  \\
Ref.~\cite{Roberts:2007ni} \footnotemark[1]
&2.492
&2.492  &2.763 &$\cdots$ &$\cdots$      
&2.592  &2.859 &$\cdots$ &$\cdots$
&2.455  &2.748 &$\cdots$ &2.768       
&2.268  &2.625 &$\cdots$ &2.816 \\
Ref.~\cite{Ebert:2011kk}
&2.476
& 2.476 & 2.792 &$\cdots$&$\cdots$     
& 2.579 & 2.854 &$\cdots$& 2.936 
& 2.443 & 2.713 &$\cdots$& 2.799 
& 2.286 & 2.598 &$\cdots$& $\cdots$ \\
Ref.~\cite{Shah:2016mig} \footnotemark[2]
&2.467
&2.470  &2.796 &$\cdots$ &2.803      
&$\cdots$ &$\cdots$ &$\cdots$ &$\cdots$
&2.452  &2.849 &$\cdots$ &2.863       
&2.286  &2.629 &$\cdots$ &$\cdots$ \\
Ref.~\cite{Garcia-Tecocoatzi:2022zrf} \footnotemark[3]
&2.466
&2.466  &2.788 &2.935 &2.977      
&2.571  &2.893 &3.040 &2.935
&2.456  &2.811 &2.994 &2.853       
&2.261  &2.616 &2.799 &2.841 \\
Ref.~\cite{Ortiz-Pacheco:2023kjn}
&2.474
&2.474  &2.777 &2.917 &2.950      
&2.586  &2.890 &3.029 &2.923
&2.455  &2.789 &2.961 &2.822       
&2.281  &2.616 &2.788 &2.821 \\
Ref.~\cite{Bijker:2020tns}
&2.461
&2.461  &2.797 &2.951 &2.980      
&2.570  &2.905 &3.060 &2.934
&$\cdots$  &$\cdots$ &$\cdots$ &$\cdots$       
&$\cdots$  &$\cdots$ &$\cdots$ &$\cdots$ \\
Used
&2.468
&2.470  &2.788  &2.935  &2.977 
&2.579  &2.893  &3.040  &2.935 
&2.453  &2.811  &2.944  &2.853 
&2.286  &2.616  &2.799  &2.841 \\
\end{tabular}%
\footnotetext[1]{~Only the unmixed data are used.}
\footnotetext[2]{~The mass that $\nu=1$ and with first-order correction.}
\footnotetext[3]{~The mass that three-quark predicted.}
\end{ruledtabular}
\label{tab:Cmass}%
\end{table}

\subsection{Numerical results and discussion}

All the Feynman diagrams for the $\Xi_c^+\to \Xi^0\pi^+$, $\Xi_c^0\to \Xi^-\pi^+$, $\Xi_c^0\to \Xi^0\pi^0$, $\Xi_c^0\to \Xi^0\eta^{(')}$ and $\Xi_c^0\to \Xi^+K^-$ processes are presented in App.~\ref{app:figamp}. These diagrams clearly show that the direct meson emission process, the color suppressed process and the pole terms (i.e. quark internal conversion process) contributions do differ.

These six channels can be classified as two categories, depending on whether the pseudoscalar meson is neutral or charged. For the neutral one, the direct meson emission process is ignored because the flavor changing neutral current (FCNC) interaction is highly suppressed in the standard model. In addition, the direct meson emission process is also forbidden in $\Xi_c^0\to \Sigma^+ K^-$, because of the charge conservation at quark level. In the quark-level transitions, one should note that the color suppressed process is absent for the $\Xi_c^0\to \Xi^- \pi^+$ and $\Xi_c^0\to \Sigma^+ K^-$ processes.

For the pole term, 
as shown by the Feynman diagrams in App.~\ref{app:figamp}, 
both the $\Xi_c^+ \to \Xi^0\pi^+$ and the $\Xi_c^0 \to \Xi^-\pi^+$ processes have only one type of pole term. In contrast, there are more pole terms contributing to the $\Xi_c^0 \to \Xi^0\eta^{(')}$ process, which is because that the constituent quarks of $\eta^{(')}$ are $u \bar u$, $d \bar d$ and $s \bar s$, making more processes are allowed. Note that the constituent quarks of $\pi^0$ are $u \bar u$ and $d \bar d$. Thus, the decay modes of the $\Xi^0\to \Xi^0\pi^0$ process also contribute to the $\Xi_c^0 \to \Xi^0\eta^{(')}$ process via the coupled channels. Although the amplitudes of each decay mode has the same form for all the processes, there is no trivial relationship between these neutral channels, due to the difference of the wave function and the relevant parameters. 

We also note that some of the intermediate states have no contributions because of the constraints from the symmetry selection rules as discussed in Ref.~\cite{Niu:2021qcc}. For example, if the  spin wave function of the initial and final states are the same, the spin matrix elements of quark-meson interaction will be vanish. This is the reasons why the $\Xi_c^0$, $|\Xi_c^0,^2P_\lambda\rangle$ and $|\Xi_c^{'0},^2P_\rho\rangle$ have no contribution to the pole term process $\Xi^+_c\to \Xi_c^0 \pi^+$.

\begin{figure}[htbp!]
\begin{center}
\includegraphics[scale=0.55]{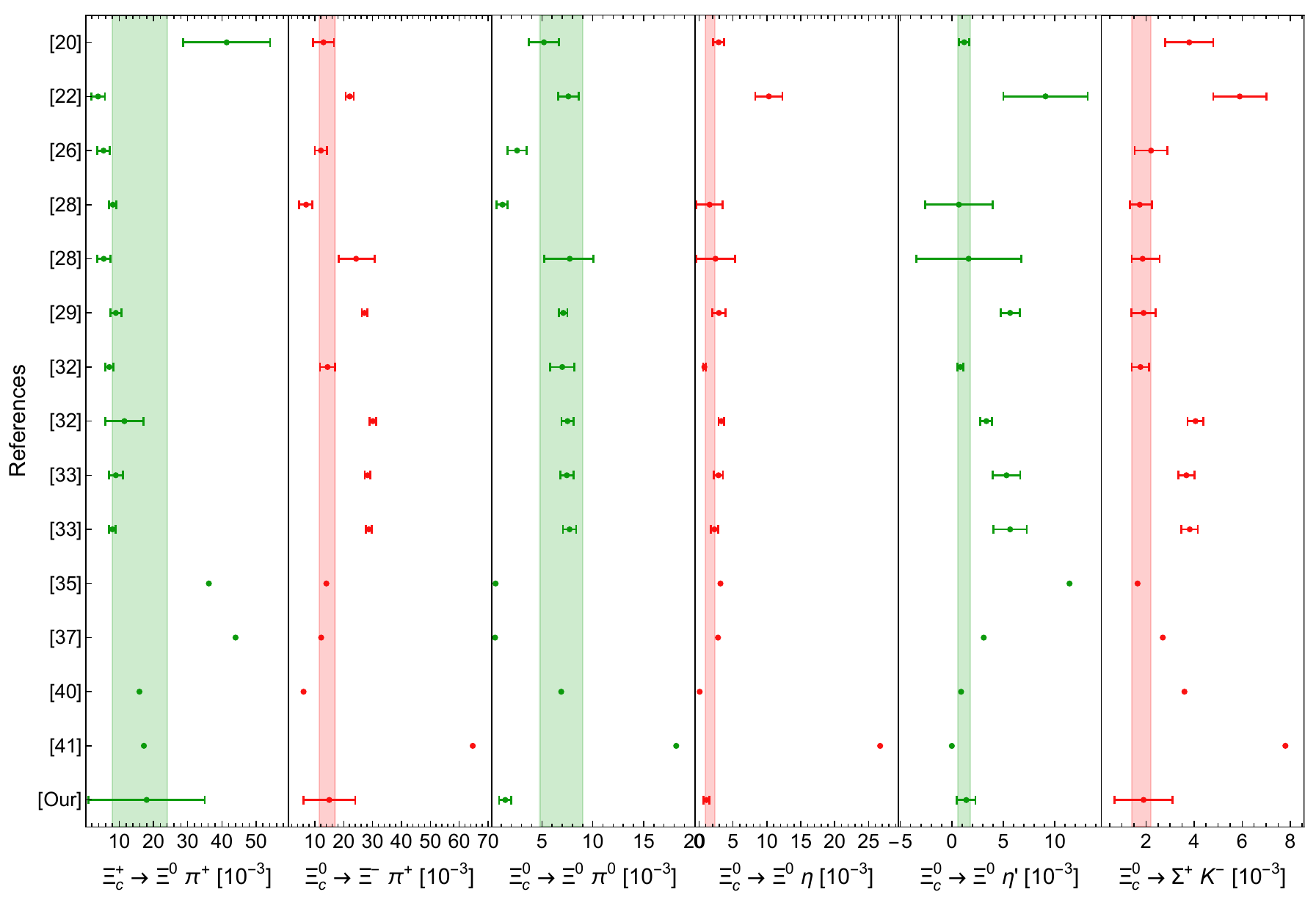}
\caption{The experimental measurements and theoretical results for the branching ratios of the six $\Xi_c^{+/0}$ decay channels. The band are the experimental region extracted from Ref.~\cite{Belle-II:2024jql} and PDG~\cite{ParticleDataGroup:2022pth}. The vertical coordinate is used to indicate the references in Tab.~\ref{tab:branch}.}
\label{fig:br}
\end{center}
\end{figure}

\begin{figure}[htbp!]
\begin{center}
\includegraphics[scale=0.55]{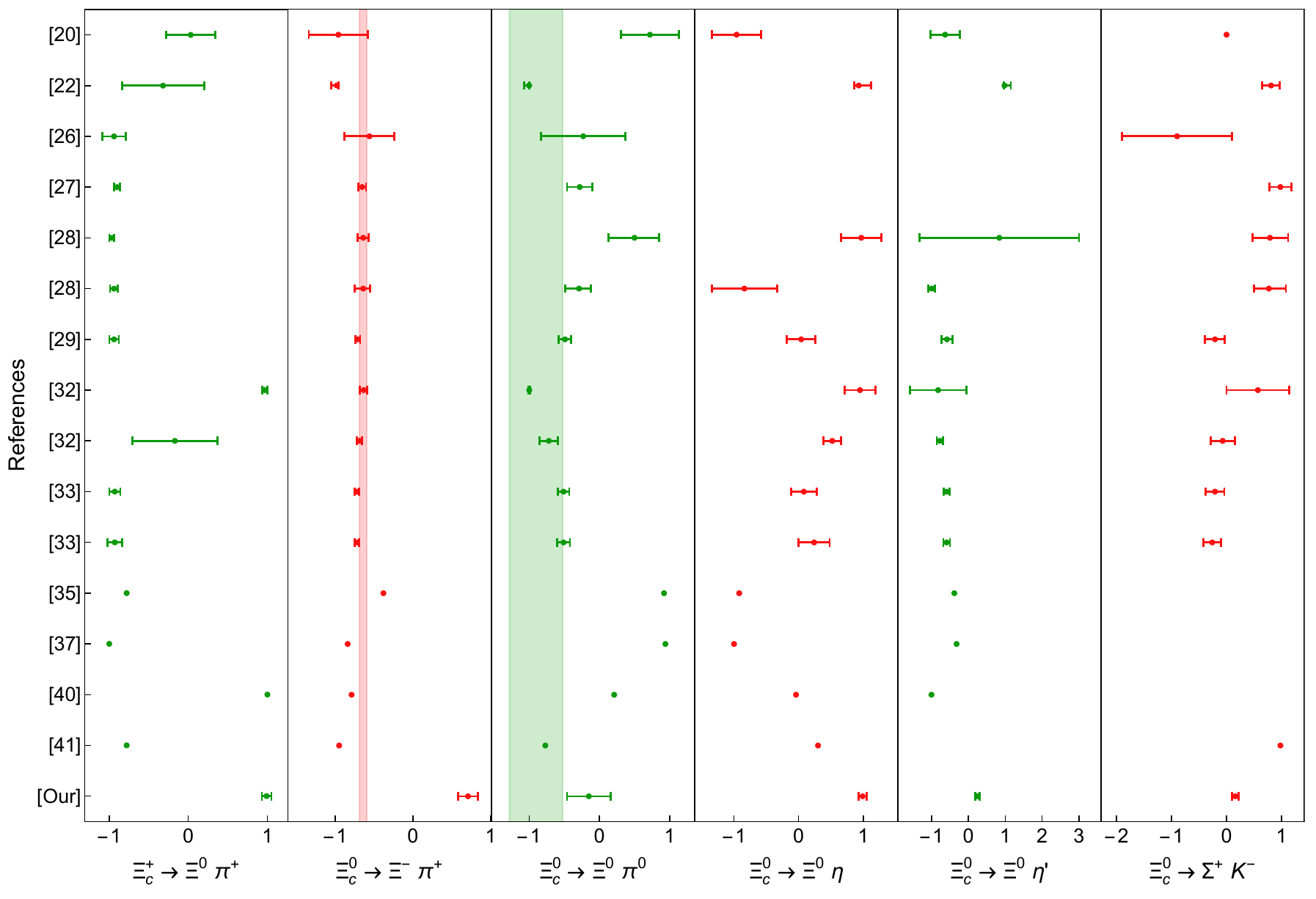}
\caption{The experimental measurements and theoretical results for the asymmetry parameters of the six $\Xi_c^{+/0}$ decay channels. The band are the experimental region extracted from Ref.~\cite{Belle-II:2024jql} and PDG~\cite{ParticleDataGroup:2022pth}. The vertical coordinate is used to indicate the references in Tab.~\ref{tab:branch}.}
\label{fig:alpha}
\end{center}
\end{figure}

The analytical forms of the transition amplitudes are tedious. Thus, only the numerical values of these amplitudes are presented in this work. The numerical details can be found in App.~\ref{app:figamp}. Figs.~\ref{fig:br} and  \ref{fig:alpha} present the branching ratios and parity asymmetry parameters of the interested channels, in comparison with other works. The detailed numbers can be found in  Tabs.~\ref{tab:branch} and \ref{tab:alpha}, respectively. 
Note that only the parity asymmetry parameters of the $\Xi_c^0\to\Xi^-\pi^+$ and the $\Xi_c^0\to \Xi^0\pi^0$ processes have been measured. The asymmetry parameters of other processes in Tab.~\ref{tab:alpha} can be regarded as a prediction.

From Fig.~\ref{fig:br}, one can see that most of our calculated branching ratios are consistent with the experimental measurements, which suggests that the non-factorizable terms play an essential role for the $\Xi_c$ hadronic weak decay. 
The branching ratio of the $\Xi_c^0\to \Xi^0 \pi^0$ process is about two sigma deviation from the experimental data. From the numerical values of the amplitudes (Tab.~\ref{tab:amp3}), one can see that the parity violating processes have little contributions due to the destructive interference among the parity violating diagrams.
This also leads to a small value for the asymmetry parameter than the experimental data. In Refs.~\cite{Han:2021azw,Jia:2024pyb} it was shown that the final states interaction may provide an additional mechanism for understanding the weak decay of $\Xi_c^0\to \Xi^0 \pi^0$. More specifically, 
the coupled channel effects between the $\Xi^0 \pi^0$ channel and the $\Xi^- \pi^+$, $\Xi^0\eta$, $\Xi^0\eta^\prime$ channels may contribute to the $\Xi_c^0\to \Xi^0 \pi^0$ branching ratio.  
On the other hand, the parity violating processes may be underestimated, for missing or overestimating the contribution of the excited charmed baryons, that may be also the reason that we obtain an opposite sign for the asymmetry parameter of the $\Xi_c^0\to \Xi^- \pi^+$ process.

In addition, we also estimate the uncertainties with the branching ratios and asymmetry parameters caused by the quark mass $m_{q/s/c}$, spring constants $K$, $R$ and the masses of the first excited charmed baryons and the results are presented in Tab.~\ref{tab:brerr} and Tab.~\ref{tab:alperr}, respectively. 
The influence of the quark masses $m_{q/s/c}$ is calculated separately. The results indicate that the transition amplitudes are sensitive to the quark masses. This is reasonable as the spatial wave function and all the operators employed in our framework rely on the quark mass. The uncertainties of the branching ratios caused by the spring constants $K$ are insignificant, because the variation of $K$ only leads a slight variation of $\alpha_\rho$ or $\alpha_\lambda$. In contrast, the direct meson emission processes are more sensitive to the parameter $R$ which has been elucidated in Ref.~\cite{Niu:2020gjw}.

Furthermore, the masses of the first excited charmed baryons also bring small uncertainty to the branching ratios. 
The excited charmed baryons only contribute to the pole terms,
whose amplitudes are proportional to the propagator of the intermediate baryon
\begin{align}
\frac{2m}{p^2-m^2} ,
\end{align}
where $m$ is the mass of the intermediate baryon. The four momentum square $p^2$ equals to $m_{B_i}^2$ or $m_{B_f}^2$,
 which depends on the types of pole terms. As the result, the uncertainty caused by $m$ can be estimated as
\begin{align}
\left|\frac{\Delta m}{m}\frac{p^2+m^2}{p^2-m^2}\right|.
\end{align}
The uncertainty from the value $m$ is about $(15\sim 20)\%$.
Nevertheless, the total uncertainty of the amplitudes caused by the uncertainties of the first excited charmed baryons is not sizeable because the contributions of these processes are almost canceled by each other, especially, in $\Xi_c^0\to \Xi^0 \pi^0$.

\begin{table}[htbp!]
\caption{The uncertainties of the branching ratios (in unit of $10^{-3}$) and the asymmetry parameters caused by $m_{q/s/c}$, $K$ and $R$ with $20\%$ and the mass of first excited charmed baryons with $10\%$. The center values and uncertainties of these parameters are listed in Tabs. \ref{tab:quarkMass}, \ref{tab:K}, \ref{tab:Lmass} and \ref{tab:Cmass}.}
\begin{ruledtabular}
\begin{tabular}{lcccccc}
Para.  &$\Xi_c^+ \to \Xi^0 \pi^+$ &  $\Xi_c^0 \to \Xi^- \pi^+$ & $\Xi_c^0 \to \Xi^0 \pi^0$ & $\Xi_c^0 \to \Xi^0 \eta$ & $\Xi_c^0 \to \Xi^0 \eta'$ & $\Xi_c^0 \to \Sigma^+ K^-$   \tabularnewline
\hline
 $m_q$&
$18\pm9$     &$15\pm4$ &$1.4\pm 0.31$ &$1.1\pm0.013$ &$1.4\pm 0.16$  &$1.9\pm0.7$ \tabularnewline
$m_s$&
$18\pm1.2$ &$15\pm1.1$ &$1.4\pm 0.15$ &$1.1\pm0.14$ &$1.4\pm 0.5$  &$1.9\pm0.7$  \tabularnewline
$m_c$&
$18\pm1.1$ &$15\pm0.4$ &$1.4\pm 0.08$ &$1.1\pm0.08$ &$1.4\pm 0.035$  &$1.9\pm0.28$  \tabularnewline
$K$ &
$18\pm0.6$ &$15\pm1.2$ &$1.4\pm 0.5$ &$1.1\pm0.4$ &$1.4\pm 0.7$  &$1.9\pm0.6$  \tabularnewline
$R$ &
$18\pm14$ &$15\pm8$ &$1.4\pm 0.15$ &$1.1\pm0.035$ &$1.4\pm 0.029$  &$1.9\pm0$  \tabularnewline
$m_{B_c^*}$ &
$18\pm1.3$ &$15\pm0$ &$1.4\pm 0.016$ &$1.1\pm0.14$ &$1.4\pm 0.20$  &$1.9\pm0$  \tabularnewline
Combined  &
$18\pm17$ &$15\pm9$ &$1.4\pm 0.6$ &$1.1\pm0.4$ &$1.4\pm 0.9$  &$1.9\pm1.2$  \tabularnewline
\end{tabular}%
\end{ruledtabular}
\label{tab:brerr}
\end{table}

\begin{table}[htbp!]
\caption{The uncertainties of the asymmetry parameters caused by $m_{q/s/c}$, $K$ and $R$ with $20\%$ and the mass of first excited charmed baryons with $10\%$. The center values and uncertainties of these parameters are listed in Tabs. \ref{tab:quarkMass}, \ref{tab:K}, \ref{tab:Lmass} and \ref{tab:Cmass}.}
\begin{ruledtabular}
\begin{tabular}{lcccccc}
Para.  &$\Xi_c^+ \to \Xi^0 \pi^+$ &  $\Xi_c^0 \to \Xi^- \pi^+$ & $\Xi_c^0 \to \Xi^0 \pi^0$ & $\Xi_c^0 \to \Xi^0 \eta$ & $\Xi_c^0 \to \Xi^0 \eta'$ & $\Xi_c^0 \to \Sigma^+ K^-$   \tabularnewline
\hline
$m_q$&
$0.99\pm0.05$ &$0.71\pm0.08$ &$-0.16\pm0.08$ &$0.99\pm0.019$ &$0.22\pm0.019$  &$0.17\pm0.020$ \tabularnewline
$m_s$&
$0.99\pm0.008$ &$0.71\pm0.033$ &$-0.16\pm 0.05$ &$0.99\pm0.010$ &$0.22\pm 0.002$  &$0.17\pm0.019$  \tabularnewline
$m_c$&
$0.99\pm0.006$ &$0.71\pm0.0008$ &$-0.16\pm 0.01$ &$0.99\pm0.007$ &$0.22\pm 0.034$  &$0.17\pm0.009$ \tabularnewline
$K$ &
$0.99\pm0.017$ &$0.71\pm0.07$ &$-0.16\pm 0.23$ &$0.99\pm0.05$ &$0.22\pm 0.04$  &$0.17\pm0.05$ \tabularnewline
$R$ &
$0.99\pm0.024$ &$0.71\pm0.05$ &$-0.16\pm 0.08$ &$0.99\pm0.004$ &$0.22\pm 0.010$  &$0.17$ \tabularnewline
$m_{B_c^*}$ &
$0.99\pm0.013$ &$0.71$ &$-0.16\pm 0.15$ &$0.99\pm0.021$ &$0.22\pm 0.018$  &$0.17$ \tabularnewline
Combined  &
$0.99\pm0.06$ &$0.71\pm0.13$ &$-0.16\pm 0.31$ &$0.99\pm0.06$ &$0.22\pm 0.06$  &$0.17\pm0.06$   \tabularnewline
\end{tabular}%
\end{ruledtabular}
\label{tab:alperr}
\end{table}

\section{Summary and Outlook}
The Cabibbo-Favored hadronic weak decay of the charmed baryon $\Xi_c$ are studied in the non-relativistic constituent quark model. Considering the availablity of the experimental data, we focus on the $\Xi_c^+\to \Xi^0\pi^+$, $\Xi_c^0\to \Xi^-\pi^+$, $\Xi_c^0\to \Xi^0\pi^0$, $\Xi_c^0\to \Xi^0\eta^{(')}$ and $\Xi_c^0\to \Xi^+K^-$ six channels. 
The contributions of direct meson emission diagrams, color suppressed diagrams and pole terms are calculated separately. The contribution of each part is channel-dependent. For example, only pole terms are allowed for the $\Xi_c^0\to \Sigma^+ K^-$ process. We employ the parameters used in the constitute quark model and estimate the branching ratios and parity asymmetry parameters of the interested six channels. 
Most of the branching ratios are consistent with the experimental data, which indicates both the factorizable diagrams (direct meson emission diagram) and non-factorizable diagrams (color suppressed diagrams and pole terms) are important 
to understand the two-body hadronic weak decay. 

As the decay asymmetry parameter is a vital physical quantity for the constraints of the dynamical mechanisms, 
we also calculate the decay asymmetry parameters of these six channels. The decay asymmetry parameters from various theoretical models vary significantly, not only the absolute value but also the relative sign. 

Although we can describe most of these quantities, We fail in obtaining the correct sign for the decay asymmetry parameter of $\Xi_c^0\to \Xi^- \pi^+$. Possible mechanisms or origin of the uncertainties are discussed. The underlying mechanism should be further investigated in both experiment and theory.

\begin{acknowledgments}
This work is partly supported by the National Natural Science Foundation of China with Grant No.~12375073,  No.~12035007, No. 12147128, and 
No. 12235018, Guangdong Provincial funding with Grant No.~2019QN01X172, Guangdong Major Project of Basic and Applied Basic Research No.~2020B0301030008, National Key Basic Research Program of China under
Contract No. 2020YFA0406300, and Strategic Priority
Research Program of Chinese Academy of Sciences
(Grant No. XDB34030302).
\end{acknowledgments}

\begin{appendix}
\section{The wave function of hadrons}
\label{app:wavefunction}

The total wave functions of baryons and pseudoscalar mesons used in calculation are presented in this section.

\subsection{The spin wave functions of hadrons}

The spin wave function is labeled as  $\chi^{\mathrm{\sigma}}_{s,s_z}$, where $\mathrm{\sigma}=\rho,\lambda, s , a$ is used to label the symmetry type of the spin wave functions which are the representations of the dimension-3 permutation group ($S_3$). The spin wave functions for the three quark system are:
\begin{align}
\chi^{\rho}_{\frac{1}{2},\frac{1}{2}}&=\frac{1}{\sqrt2}\left(\uparrow \downarrow \uparrow -\downarrow \uparrow \uparrow \right) ,
\quad& \chi^{\lambda}_{\frac{1}{2},\frac{1}{2}}&=-\frac{1}{\sqrt6}\left( \uparrow\downarrow\uparrow+\downarrow\uparrow\uparrow-2\uparrow\uparrow\downarrow \right) , \\
\chi^{\rho}_{\frac{1}{2},-\frac{1}{2}}&=\frac{1}{\sqrt2}\left(\uparrow \downarrow \downarrow -\downarrow \uparrow \downarrow \right) ,
&\chi^{\lambda}_{\frac{1}{2},-\frac{1}{2}}&=\frac{1}{\sqrt6}\left( \uparrow\downarrow\downarrow+\downarrow\uparrow\downarrow-2\downarrow\downarrow\uparrow\right) ,\\
\chi^{s}_{\frac{3}{2},\frac{3}{2}}&=\uparrow \uparrow \uparrow ,
\quad& \chi^{s}_{\frac{3}{2},-\frac{3}{2}}&= \downarrow\downarrow\downarrow  , \\
\chi^{s}_{\frac{3}{2},\frac{1}{2}}&=\frac{1}{\sqrt3}\left(\uparrow \uparrow \downarrow +\uparrow \downarrow \uparrow +\downarrow \uparrow \uparrow \right) ,
&\chi^{s}_{\frac{3}{2},-\frac{1}{2}}&=\frac{1}{\sqrt3}\left( \uparrow\downarrow\downarrow+\downarrow\uparrow\downarrow+\downarrow\downarrow\uparrow\right)  .
\end{align}
The spin wave functions for the quark-antiquark system are
\begin{align}
\chi^{s}_{1,1}=\uparrow \uparrow,\quad \chi^{s}_{1,0}=\frac{1}{\sqrt 2}\left(\uparrow \downarrow +\downarrow \uparrow\right),\quad \chi^{s}_{1,-1}=\downarrow \downarrow, \quad 
\chi^{a}_{0,0}=\frac{1}{\sqrt 2}\left(\uparrow \downarrow - \downarrow \uparrow\right).
\end{align}

\subsection{The spatial wave functions of hadrons}

The spatial wave functions are obtained by solving the Schr\"odinger equation with the harmonic oscillator potential. The Hamiltonian and more details can be found in \cite{Isgur:1978xj, LeYaouanc:1988fx, Niu:2020gjw}. In this appendix, only the spatial wave function are listed. In the momentum space, the spatial wave function of the three baryon system reads as
\begin{align}
\Psi_{N L L_z}(\bm P,\bm p_\rho,\bm p_\lambda)
=\delta^3(\bm P-\bm P_c)\sum_m \langle l_\rho,m;l_\lambda,L_z-m|L,L_z \rangle
\psi^{\alpha_\rho}_{n_\rho l_\rho m }(\bm p_\rho)
\psi^{\alpha_\lambda}_{n_\lambda l_\lambda L_z-m }(\bm p_\lambda),
\end{align}
where
\begin{align}
\psi^\alpha_{nlm}(\bm p)=(i)^l(-1)^n \left[\frac{2n!}{(n+l+1/2)!} \right]^{1/2}\frac{1}{\alpha^{l+3/2}}\exp\left({-\frac{\bm p^2}{2\alpha^2}}\right)L_n^{l+1/2}(\bm p^2/\alpha^2)
\mathcal{Y}_{lm}(\bm p) \ ,
\end{align}
where $\bm P$ is the baryon momentum; $\bm p_\rho $ and $\bm p_\lambda $ are the momentum defined in the Jacobi coordinate. $\alpha_\rho$ and $\alpha_\lambda$ are the harmonic oscillator strengths, of which the values depend on the quark mass and the spring constant $K$ (which  describes the strength of two quarks interaction) as shown in Ref.~\cite{Niu:2020gjw}.

For the light mesons, only the spatial wave function of the ground state is involved and expressed as:
\begin{equation}
\Psi_{0,0,0}(\bm p_1, \bm p_2)= \frac{1}{\pi^{3/4} R^{3/2}}\exp\left[-\frac{(\bm p_1-\bm p_2)^2}{8 R^2}\right],
\end{equation}
where $R$ is the parameter in the meson wave function, which characterizes its size.
$\bm p_1, \bm p_2$ are the three momentum of the quark and antiquark, respectively.

\subsection{Flavor wave function and total wave function of hadrons}
\subsubsection{light baryons}
The flavor wave functions for the light baryons, i.e. SU(3) flavor octet baryons~\cite{LeYaouanc:1988fx}, are:
\begin{align}
\phi_{\Lambda}^\lambda&=-\frac{1}{2}(sud+usd-sdu-dsu)  ,
\quad&\phi_{\Lambda}^\rho&=\frac{1}{2\sqrt3}(usd+sdu-sud-dsu-2dus+2uds) , \\
\phi_{\Sigma^+}^\lambda &= \frac{1}{\sqrt6}(2uus-suu-usu) ,
&\phi_{\Sigma^+}^\rho &=\frac{1}{\sqrt2}(suu-usu) ,\\
\phi_{\Sigma^0}^\lambda&=\frac{1}{2\sqrt3}(sdu+sud+usd+dsu-2uds-2dus) ,
&\phi_{\Sigma^0}^\rho&=\frac{1}{2}(sud+sdu-usd-dsu) ,\\
\phi_{\Sigma^-}^\lambda&=\frac{1}{\sqrt6}(sdd+dsd-2dds) ,
&\phi_{\Sigma^-}^\rho&=\frac{1}{\sqrt 2}(sdd-dsd) ,\\
\phi_{\Xi^0}^\lambda&=\frac{1}{\sqrt6}(2ssu-sus-uss) ,
&\phi_{\Xi^0}^\rho&=\frac{1}{\sqrt2}(sus-uss) ,\\
\phi_{\Xi^-}^\lambda&=\frac{1}{\sqrt6}(2ssd-sds-dss) ,
&\phi_{\Xi^-}^\rho&=\frac{1}{\sqrt2}(sds-dss) .
\end{align}

With the above spin, flavor and spatial parts, we can construct the total wave functions of the baryons, which are denoted by $|B, \prescript{2S+1}{}{L},J^P \rangle$. For the light baryons, the ground state reads
\begin{align}
|B, \prescript{2}{}{S}\half^+ \rangle=\frac{1}{\sqrt2}(\phi_B^\rho \chi^\rho_{S,S_z}+\phi_B^\lambda \chi^\lambda_{S,S_z})\Psi_{0,0,0} \ .
\label{eq:og}
\end{align}
For the first orbital excitation baryons,
there are two different modes, i.e. $\rho$ and  $\lambda$ configurations.
They are combined together to form the wave function for the excited baryons
\begin{align}
|B, \prescript{2}{}{P}\half^- \rangle &=\sum_{L_z+S_z=J_z}\langle1, L_z;\half, S_z|J J_z\rangle\frac{1}{2}\left[(\phi_B^\rho \chi_{S,S_z}^\lambda+\phi_B^\lambda \chi_{S,S_z}^\rho)\Psi^\rho_{1,1,L_z}+(\phi_B^\rho \chi_{S,S_z}^\rho-\phi_B^\lambda \chi_{S,S_z}^\lambda)\Psi^\lambda_{1,1,L_z} \right] , \\
|B, \prescript{4}{}{P}\half^- \rangle &=\sum_{L_z+S_z=J_z}\langle1, L_z;\frac{3}{2}, S_z|J J_z\rangle\frac{1}{\sqrt 2}\left[\phi_B^\rho \chi_{S,S_z}^s \Psi^\rho_{1,1,L_z}+ \phi_B^\lambda \chi_{S,S_z}^s\Psi^\lambda_{1,1,L_z} \right] .
\label{eq:of}
\end{align}

\subsubsection{charmed baryons}

Usually, for the charmed baryons, the flavor wave function can be constructed by symmetrizing the light quarks~\cite{Copley:1979wj,Zhong:2007gp,Wang:2017kfr} and written as:
\begin{align}
\phi^{\bar 3}_c=
\begin{dcases}
\frac{1}{\sqrt2}(ud-du)c, &~ \text{for}~ \Lambda_c^+,  \\  
\frac{1}{\sqrt2}(us-su)c, &~ \text{for}~ \Xi_c^+,\\  
\frac{1}{\sqrt2}(ds-sd)c, &~ \text{for}~ \Xi_c^0,    
\end{dcases}
\quad 
\phi^{6}_c=
\begin{dcases}
uuc, &~ \text{for}~ \Sigma_c^{++},  \\  
\frac{1}{\sqrt2}(ud+du)c, &~ \text{for}~ \Sigma_c^{+},\\  
ddc, &~ \text{for}~ \Sigma_c^0, \\
\frac{1}{\sqrt2}(us+su)c, &~ \text{for}~ \Xi_c^{\prime +},\\ 
\frac{1}{\sqrt2}(ds+sd)c, &~ \text{for}~ \Xi_c^{\prime 0},\\
ssc, &~ \text{for}~ \Omega_c^0.     
\end{dcases}
\end{align}
Then, the total wave functions of the ground states and orbital excitation states can be expressed as:
\begin{align}
\Psi_{\bar 3}=\begin{dcases}
\left |^2 S\frac{1}{2}^+ \right \rangle &=\Psi^s_{00}\chi_{S_z}^\rho\phi^{\bar 3}_Q \\
\left |^2 P_\lambda\frac{1}{2}^-\right\rangle &=\Psi^\lambda_{1L_z}\chi_{S_z}^\rho\phi^{\bar 3}_Q \\
\left |^2 P_\rho\frac{1}{2}^-\right\rangle &=\Psi^\rho_{1L_z}\chi_{S_z}^\lambda\phi^{\bar 3}_Q \\
\left |^4 P_\rho\frac{1}{2}^-\right\rangle &=\Psi^\rho_{1L_z}\chi_{S_z}^s\phi^{\bar 3}_Q
\end{dcases},
\quad 
\Psi_{6}=\begin{dcases}
\left|^2 S\frac{1}{2}^+\right\rangle &=\Psi^s_{00}\chi_{S_z}^\lambda\phi^{6}_Q \\
\left|^2 P_\lambda\frac{1}{2}^-\right\rangle &=\Psi^\lambda_{1L_z}\chi_{S_z}^\lambda\phi^{6}_Q \\
\left|^2 P_\rho\frac{1}{2}^-\right\rangle &=\Psi^\rho_{1L_z}\chi_{S_z}^\rho\phi^{6}_Q \\
\left|^4 P_\lambda\frac{1}{2}^-\right\rangle &=\Psi^\lambda_{1L_z}\chi_{S_z}^s\phi^{6}_Q
\end{dcases},
\label{eq:sextet}
\end{align}
The Clebsch-Gordan series for the spin and angular-momentum addition 
\begin{align}
\left|{ }^{2 S+1} L_\sigma J^P\right\rangle=
\sum_{L_z+S_z=J_z}\left\langle L L_z, S S_z \mid J J_z\right\rangle^N \Psi_{L L_z}^\sigma \chi_{S_z} \phi_{\Lambda_c}    
\end{align}
 has been omitted for the orbital excitation states and $\sigma=\lambda,\rho,s$ that is also used to label the symmetry type of the wave
function.

\subsubsection{pseudoscalar mesons}

The flavor wave functions of $\phi_{\mathcal P} (\mathcal P=\pi^\pm,\pi^0,K^\pm)$ are written as 
\begin{align}
\phi_{\pi^+}&=u \bar d, \quad \phi_{\pi^-}=-d\bar u ,\quad \phi_{\pi^0}=-\frac{1}{\sqrt2}(u \bar u - d \bar d),\\
\phi_{K^+}&=u\bar s ,\quad \phi_{K^-}=-s \bar u.
\end{align}
The flavor function of $\eta,\eta'$ are expressed with $\phi_{\eta_q}$ and $\phi_{\eta_s}$~\cite{Feldmann:1998vh}
\begin{align}
\left(\begin{array}{c}
\phi_{\eta} \\
\phi_{\eta'}
\end{array}\right)= \left(\begin{array}{cc}
\cos\phi & -\sin \phi \\
\sin \phi& \cos\phi 
\end{array}\right) \left(\begin{array}{c}
\phi_{\eta_q} \\
\phi_{\eta_s}
\end{array}\right) \ ,
\end{align}
where 
\begin{align}
\phi_{\eta_q} =\frac{1}{\sqrt 2}(u\bar u +d \bar d), \quad 
\phi_{\eta_s} = s \bar s \ ,
\end{align}
and $\phi$ is the mixing angle. Then we have 
\begin{align}
\phi_\eta&=\frac{\cos \phi}{\sqrt 2}\left( u\bar u +d \bar d \right)-\sin \phi s\bar s
,\\
\phi_{\eta'}&=\frac{\sin \phi}{\sqrt 2}\left( u\bar u +d \bar d \right)+\cos \phi s\bar s
.
\end{align}

The total wave function of pseudoscalar mesons is written as:
\begin{align}
\Phi_{0,0,0} (\bm p_1,\bm p_2)=\delta^3(\bm p_1+\bm p_2-\bm P)\phi_p \chi^a_{0,0}\Psi_{0,0,0}(\bm p_1,\bm p_2).
\end{align}

\section{The details of the spatial wave function convolution}
\label{app:cal}
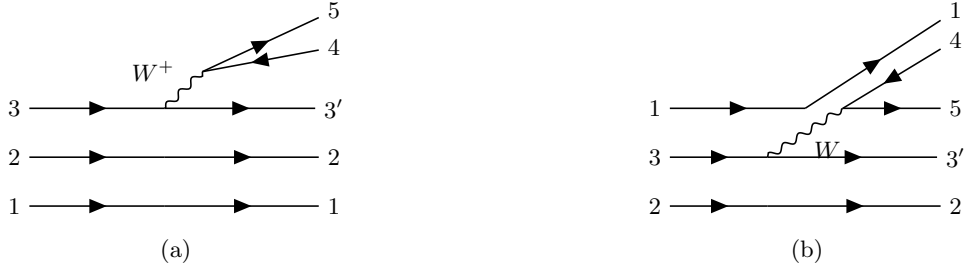
\begin{figure}[htbp!]
\begin{center}
\begin{subfigure}[h ]{0.45\textwidth}
\begin{tikzpicture}[line width=0.6pt]
\begin{feynman}
\vertex (a1) {$ 3 $};
\vertex[right=2.5cm of a1] (w1);
\vertex[right=2cm of a1] (a2);
\vertex[right=2cm of a2] (a3){$ 3' $};
\vertex[below=2em of a1] (b1) {$ 2 $};
\vertex[below=2em of a2] (b2) ;
\vertex[below=2em of a3] (b3) {$ 2 $};
\vertex[below=2em of b1] (c1) {$ 1 $};
\vertex[below=2em of b2] (c2);
\vertex[below=2em of b3] (c3) {$ 1 $};
\vertex[above=4em of a3]   (d2) {$5$};
\vertex[above=2.5em of a3] (d3) {$4$};
\vertex[above=1.5em of w1] (e1);
\diagram* {
(a1) -- [fermion] (a2) -- [fermion] (a3),
(b1) -- [fermion] (b2) -- [fermion] (b3),
(c1) -- [fermion] (c2) -- [fermion] (c3),
(e1) -- [fermion] (d2),
(d3) -- [fermion] (e1),
(a2) -- [boson, edge label=$W^{+}$] (e1),
};
\end{feynman}
\end{tikzpicture}
\caption{}
\label{fig:dpe}
\end{subfigure}
~
\begin{subfigure}[htbp! ]{0.45\textwidth}
\begin{tikzpicture}[line width=0.6pt]
\begin{feynman}
\vertex (a1) {$ 1 $};
\vertex[right=2cm of a1] (a2);
\vertex[right=0.5cm of a2](a4);
\vertex[right=4cm of a1] (a3){$ 5 $};
\vertex[below=2em of a1] (b1) {$ 3 $};
\vertex[right=1.5 of b1] (b2) ;
\vertex[below=2em of a3] (b3) {$ 3' $};
\vertex[below=2em of b1] (c1) {$ 2 $};
\vertex[right=1.5cm of c1] (c2);
\vertex[below=2em of b3] (c3) {$ 2 $};
\vertex[above=4 em of a3]  (d1) {$1$};
\vertex[above=2.8em of a3] (d2) {$4$};
\vertex[right=0.5cm of a2] (w1);
\vertex[right=1.5 of b1] (w2) ;
\diagram* {
(a1) -- [fermion] (a2),
(a4) -- [fermion] (a3),
(b1) -- [fermion] (w2) -- [fermion] (b3),
(c1) -- [fermion] (c2) -- [fermion] (c3),
(a2) -- [fermion] (d1),
(d2) -- [fermion] (w1),
(w1) -- [boson, edge label=$W$] (w2),
};
\end{feynman}
\end{tikzpicture}
\caption{}
\label{fig:cs}
\end{subfigure}\\
\end{center}
\caption{The direct pion emission (left panel) and color suppressed (right panel) Feynman diagrams in quark level. In this work the charm quark is always labeled as the third particle.}
\label{fig:label}
\end{figure}

In this work, all the integrals are performed in the momentum space.
With the quark labels are shown in Fig.~\ref{fig:label},
the general form of the spatial wave function convolution for the DME and CS processes
are written as
\begin{align}
I_{\mathrm{DME}}^{L_f,L_f^z;L_i,L_i^z}=
&\langle \psi_p(\bm k) \Psi_{N_f, L_f, L^z_f}(\bm P_f) |\hat O_{W,1\to3}^\text{spatial}(\bm p_i)|\Psi_{N_i, L_i,L^z_i}(\bm P_i)\rangle \notag \\
=&\int d \bm p_1 d \bm p_2 d \bm p_3 d \bm p_{3^\prime} d \bm p_4 d \bm p_5
\Psi_{N_f,L_f,L_f^z}^*(\bm p_1,\bm p_2,\bm p_{3^\prime})\delta^3(\bm P_f-\bm p_1-\bm p_2-\bm p_{3^\prime})  \notag \\
\times&\Psi_{0,0,0}^*(\bm p_4,\bm p_5)\delta^3(\bm k-\bm p_5-\bm p_4)  \hat O_{W,1\to3}^\text{spatial}(\bm p_i) \Psi_{N_i,L_i,L_i^z}(\bm p_1,\bm p_2,\bm p_3)\delta^3(\bm P_i-\bm p_1-\bm p_2-\bm p_3) \notag \\
\times& \delta^3(\bm p_3-\bm p_4-\bm p_5-\bm p_{3^\prime}) ,\\
I_{\mathrm{CS}}^{L_f,L_f^z; L_i,L_i^z}=&\langle \psi_p(\bm k) \Psi_{N_f, L_f, L^z_f}(\bm P_f)|\hat O_{W,1\to3}^\text{spatial}(\bm p_i)|\Psi_{N_i, L_i, L^z_i}(\bm P_i) \rangle \notag \\
=&\int d \bm p_1 d \bm p_2 d \bm p_3 d \bm p_{3^\prime} d \bm p_4 d \bm p_5
\Psi_{N_f, L_f, L_f^z}^*(\bm p_5,\bm p_2,\bm p_{3^\prime})\delta^3(\bm P_f-\bm p_5-\bm p_2-\bm p_{3^\prime}) \notag \\
\times&\Psi_{0,0,0}^*(\bm p_1,\bm p_4)\times\delta^3(\bm k-\bm p_1-\bm p_4) \hat O_{W,1\to3}^\text{spatial}(\bm p_i) \Psi_{N_i, L_i, L_i^z}(\bm p_1,\bm p_2,\bm p_3)\delta^3(\bm P_i-\bm p_1-\bm p_2-\bm p_3) \notag \\
\times&\delta^3(\bm p_3-\bm p_4-\bm p_5-\bm p_{3^\prime}) ,
\end{align}
where the operator $\hat O_{W,1\to3}^\text{spatial}(\bm p_i)$ is the function of quark momentum $\bm p_i$.

\section{The Feynman diagrams and numerical results}
\label{app:figamp}

In this section, we list all the Feynman diagrams and numerical results for the interested decay processes in this work. The data used in Figs.~\ref{fig:br} and \ref{fig:alpha} are listed in Tabs.~\ref{tab:branch} and \ref{tab:alpha}, respectively.

\begin{table}[htbp!]
\caption{The branching ratios (in unit of $10^{-3}$) shown in Fig.~\ref{fig:br}. The life time of $\tau_{\Xi_c^+}=\Xi_c^+$ ($(4.53\pm0.05)\times 10^{-13}~\mathrm{s}$) and $\tau_{\Xi_c^0}=\Xi_c^0$ ($(1.504\pm0.028)\times 10^{-13}~\mathrm{s}$) are taken from PDG~\cite{ParticleDataGroup:2022pth}.}
\begin{ruledtabular}
\begin{tabular}{lcccccc}
Model.  &$\Xi_c^+ \to \Xi^0 \pi^+$ &  $\Xi_c^0 \to \Xi^- \pi^+$ & $\Xi_c^0 \to \Xi^0 \pi^0$ & $\Xi_c^0 \to \Xi^0 \eta$ & $\Xi_c^0 \to \Xi^0 \eta'$ & $\Xi_c^0 \to \Sigma^+ K^-$   \tabularnewline
\hline
PDG~\cite{ParticleDataGroup:2022pth}   
& $16 \pm 8$      &$14.3\pm2.7$    & -             & -             & -             &$1.8\pm 0.4$    \tabularnewline
Belle\cite{Belle-II:2024jql} 
&  -              &  -             &$6.9\pm2.1$    &$1.6\pm0.7$    &$1.2\pm0.6$    & -              \tabularnewline
$\mathrm{SU}(3)_\mathrm{F}$ \cite{Sharma:1996sc}
&$41.4\pm12.7$    &$13.0\pm3.6$    &$5.2\pm1.5$    &$2.9\pm0.8$    &$1.2\pm0.5$    &$3.8\pm1.0$      \tabularnewline
$\mathrm{SU}(3)_\mathrm{F}$ \cite{Geng:2019xbo}
&$3.8\pm2.0$      &$22.1\pm1.4$    &$7.6\pm1.0$    &$10.3\pm2.0$   &$9.1\pm4.1$    &$5.9\pm1.1$      \tabularnewline
$\mathrm{SU}(3)_\mathrm{F}$ \cite{Huang:2021aqu}
&$5.4\pm1.8$      &$12.1\pm2.1$    &$2.56\pm0.93$  & -             & -             &$2.21\pm0.68$    \tabularnewline
$\mathrm{SU}(3)_\mathrm{F}$ \cite{Xing:2023dni}
&$8.87\pm0.8$     &$10.6\pm2.0$    &$1.30\pm0.51$  & $[1.93,4.46]$   &  $>0.02$      &$1.88\pm0.39$    \tabularnewline
$\mathrm{SU}(3)_\mathrm{F}$ \cite{Zhong:2022exp}
&$8.13^{+1.11}_{-1.06}$  &$6.98^{+2.48}_{-2.17}$   &$1.13^{+0.59}_{-0.49}$ & $1.56\pm1.92$          & $0.683^{+3.272}_{-3.268}$ &$1.74^{+0.41}_{-0.51}$   \tabularnewline
$\mathrm{SU}(3)_\mathrm{F}^{\mathrm{broken}}$ \cite{Zhong:2022exp}
&$5.47^{+1.95}_{-2.02}$  &$24.3^{+6.0}_{-6.4}$     &$7.74^{+2.52}_{-2.32}$ & $2.43^{+2.79}_{-2.90}$ & $1.63^{+5.09}_{-5.14}$    &$1.86^{+0.45}_{-0.71}$   \tabularnewline
$\mathrm{SU}(3)_\mathrm{F}$ \cite{Geng:2023pkr}
&$9.0\pm1.6$             &$27.2\pm0.9$            &$7.10\pm0.41$          & $2.94\pm0.97$          & $5.66\pm0.93$             &$1.9\pm0.5$             \tabularnewline
$\mathrm{SU}(3)_\mathrm{F}^{\mathrm{I}}$ \cite{Xing:2024nvg}
&$7.1\pm1.2$             &$14.4\pm2.6$             &$7\pm1.2$              & $0.79\pm0.19$          & $0.84\pm0.28$              &$1.77\pm0.36$           \tabularnewline
$\mathrm{SU}(3)_\mathrm{F}^{\mathrm{II}}$ \cite{Xing:2024nvg}
&$11.5\pm5.5$            &$30.1\pm1.2$            &$7.52\pm0.60$          & $3.30\pm0.41$           & $3.35\pm0.57$              &$4.06\pm0.33$         \tabularnewline
$\mathrm{SU}(3)_\mathrm{F}^{\mathrm{Dia}}$ \cite{Zhong:2024qqs}
&$9\pm2$                 &$28.3\pm1.0$            &$7.45\pm0.64$          & $2.87\pm0.66$          & $5.31\pm1.33$             &$3.68\pm0.34$          \tabularnewline
$\mathrm{SU}(3)_\mathrm{F}^{\mathrm{Ir}}$ \cite{Zhong:2024qqs}
&$8\pm1$                &$28.7\pm1.0$,             &$7.72\pm0.65$          & $2.28\pm0.53$          & $5.66\pm1.62$             &$3.82\pm0.34$           \tabularnewline
Quark \cite{Korner:1992wi}\footnotemark[1]
&36.2  &13.99   &0.45   &3.16  &11.43  &1.65  \tabularnewline
Quark \cite{Ivanov:1997ra}
&44.0   &12.2    &0.4    &2.8   &3.1    &2.7  \tabularnewline
Pole \cite{Zenczykowski:1993jm}
&15.9  &6.1     &6.9    &0.1   &0.9   &3.6   \tabularnewline
Pole \cite{Zou:2019kzq}\footnotemark[2]
&17.2   &64.7    &18.2   &26.7  &-      &7.8    \tabularnewline
our work  
& $18\pm17$ &$15\pm9$ &$1.4\pm 0.6$ &$1.1\pm0.4$ &$1.4\pm 0.9$  &$1.9\pm1.2$  \tabularnewline
\end{tabular}%
\footnotetext[1]{~In Ref.~\cite{Korner:1992wi} the unit of decay width is $10^11 \mathrm{s}^{-1}$.}
\footnotetext[2]{~Only the center values are listed.}
\end{ruledtabular}
\label{tab:branch}
\end{table}

\begin{table}[htbp!]
\caption{The values of the decay asymmetry parameters shown in Fig.~\ref{fig:alpha}. }
\begin{ruledtabular}
\begin{tabular}{lcccccc}
Model.  &$\Xi_c^+ \to \Xi^0 \pi^+$ &  $\Xi_c^0 \to \Xi^- \pi^+$ & $\Xi_c^0 \to \Xi^0 \pi^0$ & $\Xi_c^0 \to \Xi^0 \eta$ & $\Xi_c^0 \to \Xi^0 \eta'$ & $\Xi_c^0 \to \Sigma^+ K^-$  \tabularnewline
\hline 
PDG~\cite{ParticleDataGroup:2022pth} 
& -      &$-0.64\pm0.05$    & -               & -   & -      &-   \tabularnewline
Bell\cite{Belle-II:2024jql} 
&  -     & -                &$-0.90\pm0.38$   &-    &-       & -  \tabularnewline
$\mathrm{SU}(3)_\mathrm{F}$ \cite{Sharma:1996sc}
&$0.03\pm0.31$ &$-0.96\pm0.38$ &$0.72\pm0.41$ &$-0.96\pm0.38$ &$-0.63\pm0.40$ & 0.00   \tabularnewline
$\mathrm{SU}(3)_\mathrm{F}$ \cite{Geng:2019xbo}
&$-0.32\pm0.52$  &$-0.98^{+0.07}_{-0.02}$  &$-1.00^{+0.07}_{-0.00}$ &$0.93^{+0.07}_{-0.19}$ &$0.98^{+0.02}_{-0.17}$   &$0.81\pm0.16$   \tabularnewline
$\mathrm{SU}(3)_\mathrm{F}$ \cite{Huang:2021aqu}
&$-0.94\pm0.15$  &$-0.56\pm0.32$           &$-0.23\pm0.60$         & -                      & -                        &$-0.9\pm 1.0$    \tabularnewline
$\mathrm{SU}(3)_\mathrm{F}$ \cite{Xing:2023dni}
&$-0.902\pm0.039$&$-0.654\pm0.050$         &$-0.28\pm0.18$         &-                       &-                         &$0.98\pm 0.20$ \tabularnewline
$\mathrm{SU}(3)_\mathrm{F}$ \cite{Zhong:2022exp}
&$-0.97\pm0.03$  &$-0.64\pm0.07$           &$0.50^{+0.37}_{-0.35}$ &$0.97\pm0.31$           & $0.84\pm2.16$            &$0.79^{+0.32}_{-0.33}$  \tabularnewline
$\mathrm{SU}(3)_\mathrm{F}^{\mathrm{broken}}$ \cite{Zhong:2022exp}
&$-0.94\pm0.05$  &$-0.64^{+0.11}_{-0.09}$    &$-0.29^{+0.20}_{-0.17}$&$-0.84^{+0.50}_{-0.51}$ &$-0.99\pm0.09$            &$0.77^{+0.27}_{-0.31}$  \tabularnewline
$\mathrm{SU}(3)_\mathrm{F}$ \cite{Geng:2023pkr}
&$-0.94\pm0.06$  &$-0.71\pm0.03$           &$-0.49\pm0.09$         &$0.04\pm0.22$           &$-0.58\pm0.15$            &$-0.21\pm 0.18$  \tabularnewline

$\mathrm{SU}(3)_\mathrm{F}^{\mathrm{I}}$ \cite{Xing:2024nvg}
&$0.966\pm0.033$  &$-0.635\pm0.049$       &$-0.9982\pm0.0069$       & $0.95\pm0.24$    & $-0.82\pm0.77$              &$0.57\pm0.57$            \tabularnewline
$\mathrm{SU}(3)_\mathrm{F}^{\mathrm{II}}$ \cite{Xing:2024nvg}
&$-0.17\pm0.54$            &$-0.689\pm0.034$            &$-0.72\pm0.13$          & $0.52\pm0.14$   & $-0.775\pm0.081$              &$-0.07\pm0.22$            \tabularnewline

$\mathrm{SU}(3)_\mathrm{F}^{\mathrm{Dia}}$ \cite{Zhong:2024qqs}
&$-0.93\pm0.07$  &$-0.72\pm0.03$           &$-0.51\pm0.08$         &$0.08\pm0.20$           &$-0.59\pm0.08$            &$-0.21\pm 0.17$  \tabularnewline
$\mathrm{SU}(3)_\mathrm{F}^{\mathrm{Ir}}$ \cite{Zhong:2024qqs}
&$-0.93\pm0.09$  &$-0.72\pm0.03$           &$-0.51\pm0.09$         &$0.24\pm0.24$           &$-0.59\pm0.09$            &$-0.26\pm 0.16$  \tabularnewline
Quark \cite{Korner:1992wi}
&$-0.78$  &$-0.38$  &$0.92$  &$-0.92$  &$-0.38$   & 0   \tabularnewline
Quark \cite{Ivanov:1997ra}
&$-1.00$    &$-0.84$  &$0.94$  &$-1.0$   &$-0.32$   &$0$    \tabularnewline
Pole \cite{Zenczykowski:1993jm}
&$1.00$  &$-0.79$   &$0.21$  &$-0.04$  &$-1.00$   &$0$    \tabularnewline
Pole \cite{Zou:2019kzq}
&$-0.78$  &$-0.95$   &$-0.77$ &$ 0.30$  &-       &$0.98$ \tabularnewline
our work  &$0.99\pm0.06$ &$0.71\pm0.13$ &$-0.16\pm 0.31$ &$0.99\pm0.06$ &$0.22\pm 0.06$  &$0.17\pm0.06$   \tabularnewline
\end{tabular}
\end{ruledtabular}
\label{tab:alpha}
\end{table}

The analytical form of the transition amplitudes are tedious, so only the numerical results are provided in this work. The transition amplitudes $\mathcal M^{s^i_z,s^f_z}_{\mathrm{PC}/\mathrm{PV}}$ for the hadronic weak decay satisfy
\begin{align}
\mathcal M_{\mathrm{PC}}^{-\frac{1}{2},-\frac{1}{2}}=-\mathcal M_{\mathrm{PC}}^{\frac{1}{2},\frac{1}{2}},\quad
\mathcal M_{\mathrm{PV}}^{-\frac{1}{2},-\frac{1}{2}}=\mathcal M_{\mathrm{PV}}^{-\frac{1}{2},-\frac{1}{2}},\quad
\mathcal M^{\pm\frac{1}{2},\mp\frac{1}{2}}_{\mathrm{PC}/\mathrm{PV}}=0.
\end{align}
The superscripts $s_z^{i/f}$ are the third component of spin of the initial baryon and the finial baryon. Only the values of $\mathcal M_{\mathrm{PC}/\mathrm{PV}}^{-\frac{1}{2},-\frac{1}{2}}$ are listed in the following.

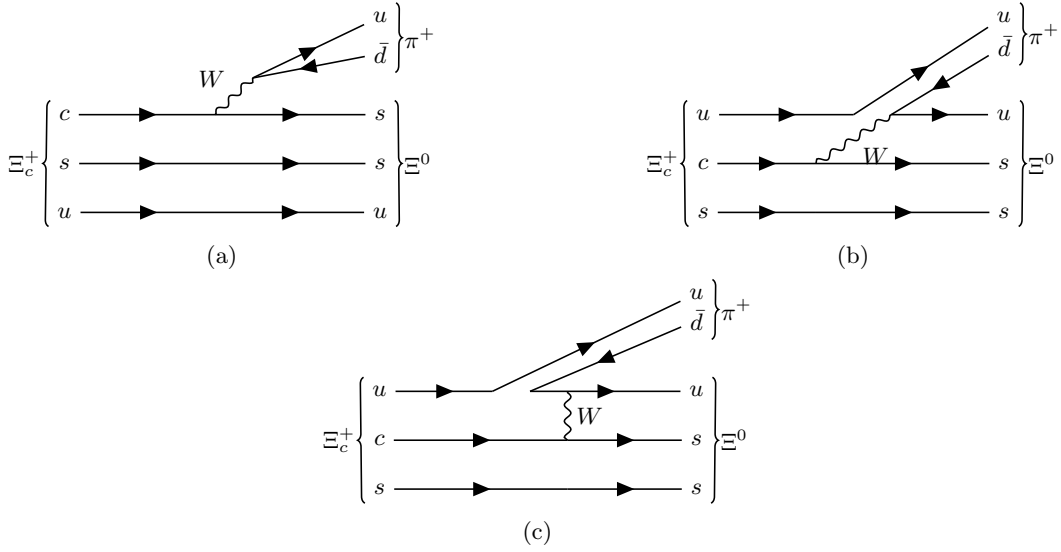
\begin{figure}[htbp!]
\centering
\begin{subfigure}[h ]{0.45\textwidth}
\begin{tikzpicture}[line width=0.6pt]
\begin{feynman}
\vertex (a1) {$ c $};
\vertex[right=2.5cm of a1] (w1);
\vertex[right=2cm of a1] (a2);
\vertex[right=2cm of a2] (a3){$ s $};
\vertex[below=2em of a1] (b1) {$ s $};
\vertex[below=2em of a2] (b2) ;
\vertex[below=2em of a3] (b3) {$ s $};
\vertex[below=2em of b1] (c1) {$ u $};
\vertex[below=2em of b2] (c2);
\vertex[below=2em of b3] (c3) {$ u $};
\vertex[above=4em of a3]   (d2) {$u$};
\vertex[above=2.5em of a3] (d3) {$\bar d$};
\vertex[above=1.5em of w1] (e1);
\diagram* {
(a1) -- [fermion] (a2) -- [fermion] (a3),
(b1) -- [fermion] (b2) -- [fermion] (b3),
(c1) -- [fermion] (c2) -- [fermion] (c3),
(e1) -- [fermion] (d2),
(d3) -- [fermion] (e1),
(a2) -- [boson, edge label=$W$] (e1),
};
\draw [decoration={brace}, decorate] (c1.south west) -- (a1.north west)
node [pos=0.5, left] {$\Xi_c^+$};
\draw [decoration={brace}, decorate] (a3.north east) -- (c3.south east)
node [pos=0.5, right] {$\Xi^0$};
\draw [decoration={brace}, decorate] (d2.north east) -- (d3.south east)
node [pos=0.5, right] {$\pi^+$};
\end{feynman}
\end{tikzpicture}
\caption{ }
\end{subfigure}
~
\begin{subfigure}[htbp! ]{0.45\textwidth}
\begin{tikzpicture}[line width=0.6pt]
\begin{feynman}
\vertex (a1) {$ u $};
\vertex[right=2cm of a1] (a2);
\vertex[right=0.5cm of a2](a4);
\vertex[right=4cm of a1] (a3){$ u $};
\vertex[below=2em of a1] (b1) {$ c $};
\vertex[right=1.5 of b1] (b2) ;
\vertex[below=2em of a3] (b3) {$ s $};
\vertex[below=2em of b1] (c1) {$ s $};
\vertex[right=1.5cm of c1] (c2);
\vertex[below=2em of b3] (c3) {$ s $};
\vertex[above=4 em of a3]  (d1) {$u$};
\vertex[above=2.8em of a3] (d2) {$\bar d$};
\vertex[right=0.5cm of a2] (w1);
\vertex[right=1.5 of b1] (w2) ;
\diagram* {
(a1) -- [fermion] (a2),
(a4) -- [fermion] (a3),
(b1) -- [fermion] (w2) -- [fermion] (b3),
(c1) -- [fermion] (c2) -- [fermion] (c3),
(a2) -- [fermion] (d1),
(d2) -- [fermion] (w1),
(w1) -- [boson, edge label=$W$] (w2),
};
\draw [decoration={brace}, decorate] (c1.south west) -- (a1.north west)
node [pos=0.5, left] {$\Xi_c^+$};
\draw [decoration={brace}, decorate] (a3.north east) -- (c3.south east)
node [pos=0.5, right] {$\Xi^0$};
\draw [decoration={brace}, decorate] (d1.north east) -- (d2.south east)
node [pos=0.5, right] {$\pi^+$};
\end{feynman}
\end{tikzpicture}
\caption{}
\end{subfigure}\\

\begin{subfigure}[htbp! ]{0.45\textwidth}
\begin{tikzpicture}[line width=0.6pt]
\begin{feynman}
\vertex (a1) {$ u $};
\vertex[right=1.5cm of a1] (a2);
\vertex[right=0.5cm of a2] (a3);
\vertex[right=0.5cm of a3] (a4);
\vertex[right=1.5cm of a4] (a5){$ u $};
\vertex[below=2em of a1] (b1) {$ c $};
\vertex[below=2em of a4] (b2) ;
\vertex[below=2em of a5] (b3) {$ s $};
\vertex[below=2em of b1] (c1) {$ s $};
\vertex[below=2em of b2] (c2);
\vertex[below=2em of b3] (c3) {$ s $};
\vertex[above=4em of a5]   (d1) {$u$};
\vertex[above=2.9em of a5] (d2) {$\bar d$};
\vertex[right=2.5cm of a1] (w1);
\vertex[below=2em of w1] (w2);
\diagram* {
(a1) -- [fermion] (a2),
(a3) -- [fermion] (a5),
(b1) -- [fermion] (b2) -- [fermion] (b3),
(c1) -- [fermion] (c2) -- [fermion] (c3),
(a2) -- [fermion] (d1),
(d2) -- [fermion] (a3),
(w1) -- [boson, edge label=$W$] (w2),
};
\draw [decoration={brace}, decorate] (c1.south west) -- (a1.north west)
node [pos=0.5, left] {$\Xi_c^+$};
\draw [decoration={brace}, decorate] (a5.north east) -- (c3.south east)
node [pos=0.5, right] {$\Xi^0$};
\draw [decoration={brace}, decorate] (d1.north east) -- (d2.south east)
node [pos=0.5, right] {$\pi^+$};
\end{feynman}
\end{tikzpicture}
\caption{}
\end{subfigure}
\caption{The transition diagrams for $\Xi_c^+\to \Xi^0\pi^+$. (a) direct meson emission process; (b) color suppressed pion emission process; (c) quark internal conversion process.
}
\label{fig:Xizpip}
\end{figure}

\begin{table}[htbp!]
\centering
\footnotesize
\caption{The amplitudes of $\Xi_c^+\to \Xi^0\pi^+$ (in unit of $10^{-9}\mathrm{GeV}^{-1/2}$ for the real part and $10^{-13}\mathrm{GeV}^{-1/2}$ for the imaginary part). WS (SW) is used to label the pole terms that baryon weak transition
either preceding (following) the strong meson emission.}
\begin{ruledtabular} 
\begin{tabular}{cllllllllll}
   & DME       &  CS     &  WS  &   \multicolumn{6}{l}{SW}   & Total \\
\hline
   & \y        &  \y     &  \n  &  $\Xi_c^0$  &           &       &  $\Xi_c^{'0}$  &    &    &    \\ 
PC & $34.41$   & $-5.62$ &  0   &  $ 0$       &           &       &  $ -8.47$      &    &    &  $20.32$  \\
\hline
   & \y        &  \y     &   \n &  $|^2P_\rho\rangle$  & $|^2P_\lambda\rangle$ & $|^4P_\rho\rangle$  & $|^2P_\rho\rangle$ & $|^2P_\lambda\rangle$ & $|^4P_\lambda\rangle$ &     \\
PV & $-18.98$  &$5.34$   &   0  &$2.84+6.04i$             &              0     & $-4.22-8.80i$       &   0                   & $4.12+8.98i$       & $-6.22-13.26i$         & $-17.11-7.03i$  \\
\end{tabular}
\end{ruledtabular}
\label{tab:amp1}%
\end{table}%

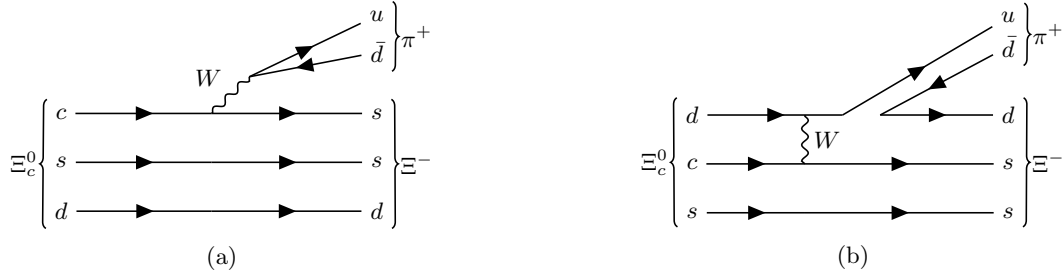
\begin{figure}[htbp!]
\centering
\begin{subfigure}[h ]{0.45\textwidth}
\begin{tikzpicture}[line width=0.6pt]
\begin{feynman}
\vertex (a1) {$ c $};
\vertex[right=2.5cm of a1] (w1);
\vertex[right=2cm of a1] (a2);
\vertex[right=2cm of a2] (a3){$ s $};
\vertex[below=2em of a1] (b1) {$ s $};
\vertex[below=2em of a2] (b2) ;
\vertex[below=2em of a3] (b3) {$ s $};
\vertex[below=2em of b1] (c1) {$ d $};
\vertex[below=2em of b2] (c2);
\vertex[below=2em of b3] (c3) {$ d $};
\vertex[above=4em of a3]   (d2) {$u$};
\vertex[above=2.5em of a3] (d3) {$\bar d$};
\vertex[above=1.5em of w1] (e1);
\diagram* {
(a1) -- [fermion] (a2) -- [fermion] (a3),
(b1) -- [fermion] (b2) -- [fermion] (b3),
(c1) -- [fermion] (c2) -- [fermion] (c3),
(e1) -- [fermion] (d2),
(d3) -- [fermion] (e1),
(a2) -- [boson, edge label=$W$] (e1),
};
\draw [decoration={brace}, decorate] (c1.south west) -- (a1.north west)
node [pos=0.5, left] {$\Xi_c^0$};
\draw [decoration={brace}, decorate] (a3.north east) -- (c3.south east)
node [pos=0.5, right] {$\Xi^-$};
\draw [decoration={brace}, decorate] (d2.north east) -- (d3.south east)
node [pos=0.5, right] {$\pi^+$};
\end{feynman}
\end{tikzpicture}
\caption{ }
\end{subfigure}
~
\begin{subfigure}[htbp! ]{0.45\textwidth}
\begin{tikzpicture}[line width=0.6pt]
\begin{feynman}
\vertex (a1) {$ d $};
\vertex[right=1.5cm of a1] (a2);
\vertex[right=0.5cm of a2] (a3);
\vertex[right=0.5cm of a3] (a4);
\vertex[right=1.5cm of a4] (a5){$ d $};
\vertex[below=2em of a1] (b1) {$ c $};
\vertex[below=2em of a2] (b2) ;
\vertex[below=2em of a5] (b3) {$ s $};
\vertex[below=2em of b1] (c1) {$ s $};
\vertex[below=2em of b2] (c2);
\vertex[below=2em of b3] (c3) {$ s $};
\vertex[above=4em of a5]   (d1) {$u$};
\vertex[above=2.8em of a5] (d2) {$\bar d$};
\diagram* {
(a1) -- [fermion] (a3),
(a4) -- [fermion] (a5),
(b1) -- [fermion] (b2) -- [fermion] (b3),
(c1) -- [fermion] (c2) -- [fermion] (c3),
(a3) -- [fermion] (d1),
(d2) -- [fermion] (a4),
(a2) -- [boson, edge label=$W$] (b2),
};
\draw [decoration={brace}, decorate] (c1.south west) -- (a1.north west)
node [pos=0.5, left] {$\Xi_c^0$};
\draw [decoration={brace}, decorate] (a5.north east) -- (c3.south east)
node [pos=0.5, right] {$\Xi^-$};
\draw [decoration={brace}, decorate] (d1.north east) -- (d2.south east)
node [pos=0.5, right] {$\pi^+$};
\end{feynman}
\end{tikzpicture}
\caption{ }
\end{subfigure}
\caption{The transition diagrams for $\Xi_c^0\to \Xi^-\pi^+$. (a) direct meson emission process; (b) quark internal conversion process.}
\label{fig:Xippip}
\end{figure}

\begin{table}[htbp!]
\centering
\footnotesize
\caption{The amplitudes of the $\Xi_c^0\to \Xi^-\pi^+$ in unit of $10^{-9}\mathrm{GeV}^{-1/2}$.}
\begin{ruledtabular} 
\begin{tabular}{cllllll}
   & DME      &  CS  & \multicolumn{2}{l}{WS}                               &     SW    & Total \\
\hline
   &  \y      &  \n  & $ \Xi_c^0$                &                          &     \n    &      \\ 
PC & $34.41$  &   0  & $3.37-3.04\times10^{-2}i$  &                          &     0     & $37.78-3.04\times10^{-2}i$  \\
\hline
   &  \y      &  \n  & $\Xi_c^{*0}$              & $\Xi_c^{0}(1690)$        &     \n    &     \\
PV & $-19.02$ &   0  & $0.36-5.37\times10^{-3}i$  & $2.92-3.04\times10^{-2}i$ &     0     & $-15.74-3.58\times 10^{-2}i$  \\
\end{tabular}
\end{ruledtabular}
\label{tab:amp2}
\end{table}

\begin{figure}[htbp!]
\centering
\begin{subfigure}[htbp! ]{0.45\textwidth}
\begin{tikzpicture}[line width=0.6pt]
\begin{feynman}
\vertex (a1) {$ d $};
\vertex[right=2cm of a1] (a2);
\vertex[right=4cm of a1] (a3){$ u $};
\vertex[below=2em of a1] (b1) {$ c $};
\vertex[right=1.5 of b1] (b2) ;
\vertex[below=2em of a3] (b3) {$ s $};
\vertex[below=2em of b1] (c1) {$ s $};
\vertex[right=1.5cm of c1] (c2);
\vertex[below=2em of b3] (c3) {$ s $};
\vertex[above=4 em of a3]  (d1) {$d$};
\vertex[above=2.8em of a3] (d2) {$\bar d$};
\vertex[right=0.5cm of a2] (w1);
\vertex[right=1.5 of b1] (w2) ;
\diagram* {
(a1) -- [fermion] (a2),
(a3) -- [fermion] (a4),
(b1) -- [fermion] (w2) -- [fermion] (b3),
(c1) -- [fermion] (c2) -- [fermion] (c3),
(a2) -- [fermion] (d1),
(d2) -- [fermion] (w1),
(w1) -- [boson, edge label=$W$] (w2),
};
\draw [decoration={brace}, decorate] (c1.south west) -- (a1.north west)
node [pos=0.5, left] {$\Xi_c^0$};
\draw [decoration={brace}, decorate] (a3.north east) -- (c3.south east)
node [pos=0.5, right] {$\Xi^0$};
\draw [decoration={brace}, decorate] (d1.north east) -- (d2.south east)
node [pos=0.5, right] {$\pi^0$};
\end{feynman}
\end{tikzpicture}
\caption{}
\end{subfigure}
~
\begin{subfigure}[htbp! ]{0.45\textwidth}
\begin{tikzpicture}[line width=0.6pt]
\begin{feynman}
\vertex (a1) {$ d $};
\vertex[right=1.5cm of a1] (a2);
\vertex[right=0.5cm of a2] (a3);
\vertex[right=0.5cm of a3] (a4);
\vertex[right=1.5cm of a4] (a5){$ u $};
\vertex[below=2em of a1] (b1) {$ c $};
\vertex[below=2em of a2] (b2) ;
\vertex[below=2em of a5] (b3) {$ s $};
\vertex[below=2em of b1] (c1) {$ s $};
\vertex[below=2em of b2] (c2);
\vertex[below=2em of b3] (c3) {$ s $};
\vertex[above=4em of a5]   (d1) {$u$};
\vertex[above=2.8em of a5] (d2) {$\bar u$};
\diagram* {
(a1) -- [fermion] (a3),
(a4) -- [fermion] (a5),
(b1) -- [fermion] (b2) -- [fermion] (b3),
(c1) -- [fermion] (c2) -- [fermion] (c3),
(a3) -- [fermion] (d1),
(d2) -- [fermion] (a4),
(a2) -- [boson, edge label=$W$] (b2),
};
\draw [decoration={brace}, decorate] (c1.south west) -- (a1.north west)
node [pos=0.5, left] {$\Xi_c^0$};
\draw [decoration={brace}, decorate] (a5.north east) -- (c3.south east)
node [pos=0.5, right] {$\Xi^0$};
\draw [decoration={brace}, decorate] (d1.north east) -- (d2.south east)
node [pos=0.5, right] {$\pi^0$};
\end{feynman}
\end{tikzpicture}
\caption{}
\end{subfigure} \\

\begin{subfigure}[htbp! ]{0.45\textwidth}
\begin{tikzpicture}[line width=0.6pt]
\begin{feynman}
\vertex (a1) {$ d $};
\vertex[right=1.5cm of a1] (a2);
\vertex[right=0.5cm of a2] (a3);
\vertex[right=0.5cm of a3] (a4);
\vertex[right=1.5cm of a4] (a5){$ u $};
\vertex[below=2em of a1] (b1) {$ c $};
\vertex[below=2em of a4] (b2) ;
\vertex[below=2em of a5] (b3) {$ s $};
\vertex[below=2em of b1] (c1) {$ s $};
\vertex[below=2em of b2] (c2);
\vertex[below=2em of b3] (c3) {$ s $};
\vertex[above=4em of a5]   (d1) {$d$};
\vertex[above=2.9em of a5] (d2) {$\bar d$};
\vertex[right=2.5cm of a1] (w1);
\vertex[below=2em of w1] (w2);
\diagram* {
(a1) -- [fermion] (a2),
(a3) -- [fermion] (a5),
(b1) -- [fermion] (b2) -- [fermion] (b3),
(c1) -- [fermion] (c2) -- [fermion] (c3),
(a2) -- [fermion] (d1),
(d2) -- [fermion] (a3),
(w1) -- [boson, edge label=$W$] (w2),
};
\draw [decoration={brace}, decorate] (c1.south west) -- (a1.north west)
node [pos=0.5, left] {$\Xi_c^0$};
\draw [decoration={brace}, decorate] (a5.north east) -- (c3.south east)
node [pos=0.5, right] {$\Xi^0$};
\draw [decoration={brace}, decorate] (d1.north east) -- (d2.south east)
node [pos=0.5, right] {$\pi^0$};
\end{feynman}
\end{tikzpicture}
\caption{}
\end{subfigure}
\caption{The transition diagrams for $\Xi_c^0\to \Xi^0\pi^0$. (a) direct meson emission process; (b) and (c) quark internal conversion process.}
\label{fig:Xizpiz}
\end{figure}
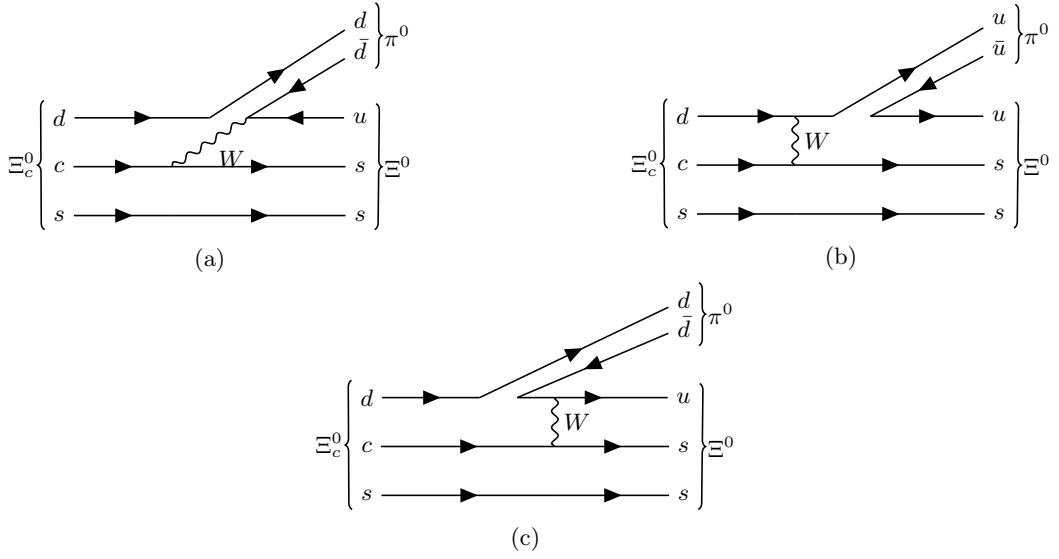

\begin{table}[htbp!]
\centering
\footnotesize
\caption{The amplitudes of $\Xi_c^0\to  \Xi^0 \pi^0$ (in unit of $10^{-9}\mathrm{GeV}^{-1/2}$ for the real part and $10^{-12}\mathrm{GeV}^{-1/2}$ for the imaginary part).}
\begin{ruledtabular} 
\begin{tabular}{clllllllllll}
   & DME &  CS    &\multicolumn{2}{l}{WS}  &   \multicolumn{6}{l}{SW}   & Total \\
\hline
   & \n  & \y     &$\Xi^0$               &                 &    $\Xi_c^0$  &           &       &  $\Xi_c^{'0}$  &    &    &    \\ 
PC & 0   & $3.98$ &$2.37-21.34i$         &                 &    $0$        &           &       &  $5.95$        &    &    & $12.29-21.34i$  \\
\hline
   & \n  &  \y    &$\Xi^{*0}$            & $\Xi^{0}(1690)$ & $|^2P_\rho\rangle$  & $|^2P_\lambda\rangle$ & $|^4P_\rho\rangle$ &  $|^2P_\rho\rangle$  & $|^2P_\lambda\rangle$ & $|^4P_\lambda\rangle$ &   \\
PV &   0 & $-3.77$& $0.25-3.66i$         & $1.99-20.73i$   & $-1.99-0.42i$          &     0              & $2.97 + 0.62i$     &         0               & $-2.90-0.63i$      & $4.38+ 0.93i$         &$0.92 - 23.90i$  \\
\end{tabular}
\end{ruledtabular}
\label{tab:amp3}%
\end{table}%

\begin{figure}[htbp!]
\centering
\begin{subfigure}[htbp! ]{0.45\textwidth}
\begin{tikzpicture}[line width=0.6pt]
\begin{feynman}
\vertex (a1) {$ d $};
\vertex[right=2cm of a1] (a2);
\vertex[right=4cm of a1] (a3){$ u $};
\vertex[below=2em of a1] (b1) {$ c $};
\vertex[right=1.5 of b1] (b2) ;
\vertex[below=2em of a3] (b3) {$ s $};
\vertex[below=2em of b1] (c1) {$ s $};
\vertex[right=1.5cm of c1] (c2);
\vertex[below=2em of b3] (c3) {$ s $};
\vertex[above=4 em of a3]  (d1) {$d$};
\vertex[above=2.8em of a3] (d2) {$\bar d$};
\vertex[right=0.5cm of a2] (w1);
\vertex[right=1.5 of b1] (w2) ;
\diagram* {
(a1) -- [fermion] (a2),
(a3) -- [fermion] (a4),
(b1) -- [fermion] (w2) -- [fermion] (b3),
(c1) -- [fermion] (c2) -- [fermion] (c3),
(a2) -- [fermion] (d1),
(d2) -- [fermion] (w1),
(w1) -- [boson, edge label=$W$] (w2),
};
\draw [decoration={brace}, decorate] (c1.south west) -- (a1.north west)
node [pos=0.5, left] {$\Xi_c^0$};
\draw [decoration={brace}, decorate] (a3.north east) -- (c3.south east)
node [pos=0.5, right] {$\Xi^0$};
\draw [decoration={brace}, decorate] (d1.north east) -- (d2.south east)
node [pos=0.5, right] {$\eta^{(\prime)}$};
\end{feynman}
\end{tikzpicture}
\caption{}
\end{subfigure}
~
\begin{subfigure}[htbp! ]{0.45\textwidth}
\begin{tikzpicture}[line width=0.6pt]
\begin{feynman}
\vertex (a1) {$ d $};
\vertex[right=1.5cm of a1] (a2);
\vertex[right=0.5cm of a2] (a3);
\vertex[right=0.5cm of a3] (a4);
\vertex[right=1.5cm of a4] (a5){$ u $};
\vertex[below=2em of a1] (b1) {$ c $};
\vertex[below=2em of a2] (b2) ;
\vertex[below=2em of a5] (b3) {$ s $};
\vertex[below=2em of b1] (c1) {$ s $};
\vertex[below=2em of b2] (c2);
\vertex[below=2em of b3] (c3) {$ s $};
\vertex[above=4em of a5]   (d1) {$u$};
\vertex[above=2.8em of a5] (d2) {$\bar u$};
\diagram* {
(a1) -- [fermion] (a3),
(a4) -- [fermion] (a5),
(b1) -- [fermion] (b2) -- [fermion] (b3),
(c1) -- [fermion] (c2) -- [fermion] (c3),
(a3) -- [fermion] (d1),
(d2) -- [fermion] (a4),
(a2) -- [boson, edge label=$W$] (b2),
};
\draw [decoration={brace}, decorate] (c1.south west) -- (a1.north west)
node [pos=0.5, left] {$\Xi_c^0$};
\draw [decoration={brace}, decorate] (a5.north east) -- (c3.south east)
node [pos=0.5, right] {$\Xi^0$};
\draw [decoration={brace}, decorate] (d1.north east) -- (d2.south east)
node [pos=0.5, right] {$\eta^{(\prime)}$};
\end{feynman}
\end{tikzpicture}
\caption{}
\end{subfigure} \\

\begin{subfigure}[htbp! ]{0.45\textwidth}
\begin{tikzpicture}[line width=0.6pt]
\begin{feynman}
\vertex (a1) {$ c $};
\vertex[right=1.5cm of a1] (a2);
\vertex[right=0.5cm of a2] (a3);
\vertex[right=0.5cm of a3] (a4);
\vertex[right=1.5cm of a4] (a5){$ s $};
\vertex[below=2em of a1] (b1) {$ d $};
\vertex[below=2em of a2] (b2) ;
\vertex[below=2em of a5] (b3) {$ u $};
\vertex[below=2em of b1] (c1) {$ s $};
\vertex[below=2em of b2] (c2);
\vertex[below=2em of b3] (c3) {$ s $};
\vertex[above=4em of a5]   (d1) {$s$};
\vertex[above=2.8em of a5] (d2) {$\bar s$};
\diagram* {
(a1) -- [fermion] (a3),
(a4) -- [fermion] (a5),
(b1) -- [fermion] (b2) -- [fermion] (b3),
(c1) -- [fermion] (c2) -- [fermion] (c3),
(a3) -- [fermion] (d1),
(d2) -- [fermion] (a4),
(a2) -- [boson, edge label=$W$] (b2),
};
\draw [decoration={brace}, decorate] (c1.south west) -- (a1.north west)
node [pos=0.5, left] {$\Xi_c^0$};
\draw [decoration={brace}, decorate] (a5.north east) -- (c3.south east)
node [pos=0.5, right] {$\Xi^0$};
\draw [decoration={brace}, decorate] (d1.north east) -- (d2.south east)
node [pos=0.5, right] {$\eta^{(\prime)}$};
\end{feynman}
\end{tikzpicture}
\caption{}
\end{subfigure} 
~
\begin{subfigure}[htbp! ]{0.45\textwidth}
\begin{tikzpicture}[line width=0.6pt]
\begin{feynman}
\vertex (a1) {$ s $};
\vertex[right=1.5cm of a1] (a2);
\vertex[right=0.5cm of a2] (a3);
\vertex[right=0.5cm of a3] (a4);
\vertex[right=1.5cm of a4] (a5){$ s $};
\vertex[below=2em of a1] (b1) {$ d $};
\vertex[below=2em of a2] (b2) ;
\vertex[below=2em of a5] (b3) {$ u $};
\vertex[below=2em of b1] (c1) {$ c $};
\vertex[below=2em of b2] (c2);
\vertex[below=2em of b3] (c3) {$ s $};
\vertex[above=4em of a5]   (d1) {$s$};
\vertex[above=2.8em of a5] (d2) {$\bar s$};
\diagram* {
(a1) -- [fermion] (a3),
(a4) -- [fermion] (a5),
(b1) -- [fermion] (b2) -- [fermion] (b3),
(c1) -- [fermion] (c2) -- [fermion] (c3),
(a3) -- [fermion] (d1),
(d2) -- [fermion] (a4),
(c2) -- [boson, edge label=$W$] (b2),
};
\draw [decoration={brace}, decorate] (c1.south west) -- (a1.north west)
node [pos=0.5, left] {$\Xi_c^0$};
\draw [decoration={brace}, decorate] (a5.north east) -- (c3.south east)
node [pos=0.5, right] {$\Xi^0$};
\draw [decoration={brace}, decorate] (d1.north east) -- (d2.south east)
node [pos=0.5, right] {$\eta^{(\prime)}$};
\end{feynman}
\end{tikzpicture}
\caption{}
\end{subfigure} 
\\
\begin{subfigure}[htbp! ]{0.45\textwidth}
\begin{tikzpicture}[line width=0.6pt]
\begin{feynman}
\vertex (a1) {$ d $};
\vertex[right=1.5cm of a1] (a2);
\vertex[right=0.5cm of a2] (a3);
\vertex[right=0.5cm of a3] (a4);
\vertex[right=1.5cm of a4] (a5){$ u $};
\vertex[below=2em of a1] (b1) {$ c $};
\vertex[below=2em of a4] (b2) ;
\vertex[below=2em of a5] (b3) {$ s $};
\vertex[below=2em of b1] (c1) {$ s $};
\vertex[below=2em of b2] (c2);
\vertex[below=2em of b3] (c3) {$ s $};
\vertex[above=4em of a5]   (d1) {$d$};
\vertex[above=2.9em of a5] (d2) {$\bar d$};
\vertex[right=2.5cm of a1] (w1);
\vertex[below=2em of w1] (w2);
\diagram* {
(a1) -- [fermion] (a2),
(a3) -- [fermion] (a5),
(b1) -- [fermion] (b2) -- [fermion] (b3),
(c1) -- [fermion] (c2) -- [fermion] (c3),
(a2) -- [fermion] (d1),
(d2) -- [fermion] (a3),
(w1) -- [boson, edge label=$W$] (w2),
};
\draw [decoration={brace}, decorate] (c1.south west) -- (a1.north west)
node [pos=0.5, left] {$\Xi_c^0$};
\draw [decoration={brace}, decorate] (a5.north east) -- (c3.south east)
node [pos=0.5, right] {$\Xi^0$};
\draw [decoration={brace}, decorate] (d1.north east) -- (d2.south east)
node [pos=0.5, right] {$\eta^{(\prime)}$};
\end{feynman}
\end{tikzpicture}
\caption{}
\end{subfigure}
~
\begin{subfigure}[htbp! ]{0.45\textwidth}
\begin{tikzpicture}[line width=0.6pt]
\begin{feynman}
\vertex (a1) {$ s $};
\vertex[right=1.5cm of a1] (a2);
\vertex[right=0.5cm of a2] (a3);
\vertex[right=0.5cm of a3] (a4);
\vertex[right=1.5cm of a4] (a5){$ s $};
\vertex[below=2em of a1] (b1) {$ d $};
\vertex[below=2em of a4] (b2) ;
\vertex[below=2em of a5] (b3) {$ u $};
\vertex[below=2em of b1] (c1) {$ c $};
\vertex[below=2em of b2] (c2);
\vertex[below=2em of b3] (c3) {$ s $};
\vertex[above=4em of a5]   (d1) {$s$};
\vertex[above=2.9em of a5] (d2) {$\bar s$};
\vertex[right=2.5cm of a1] (w1);
\vertex[below=2em of w1] (w2);
\diagram* {
(a1) -- [fermion] (a2),
(a3) -- [fermion] (a5),
(b1) -- [fermion] (b2) -- [fermion] (b3),
(c1) -- [fermion] (c2) -- [fermion] (c3),
(a2) -- [fermion] (d1),
(d2) -- [fermion] (a3),
(c2) -- [boson, edge label=$W$] (w2),
};
\draw [decoration={brace}, decorate] (c1.south west) -- (a1.north west)
node [pos=0.5, left] {$\Xi_c^0$};
\draw [decoration={brace}, decorate] (a5.north east) -- (c3.south east)
node [pos=0.5, right] {$\Xi^0$};
\draw [decoration={brace}, decorate] (d1.north east) -- (d2.south east)
node [pos=0.5, right] {$\eta^{(\prime)}$};
\end{feynman}
\end{tikzpicture}
\caption{}
\end{subfigure}
\caption{The transition diagrams of $\Xi_c^0\to  \Xi^0 \eta^{(\prime)}$. (a) color suppressed process; (b)-(f) quark internal conversion process. Considering the existence of the intermediate baryon, (c) and (d) can contribute to the same intermediate baryon. Thus we treat their sum as the amplitude from one intermediate baryon in the calculation. }
\label{fig:Xizeta}
\end{figure}
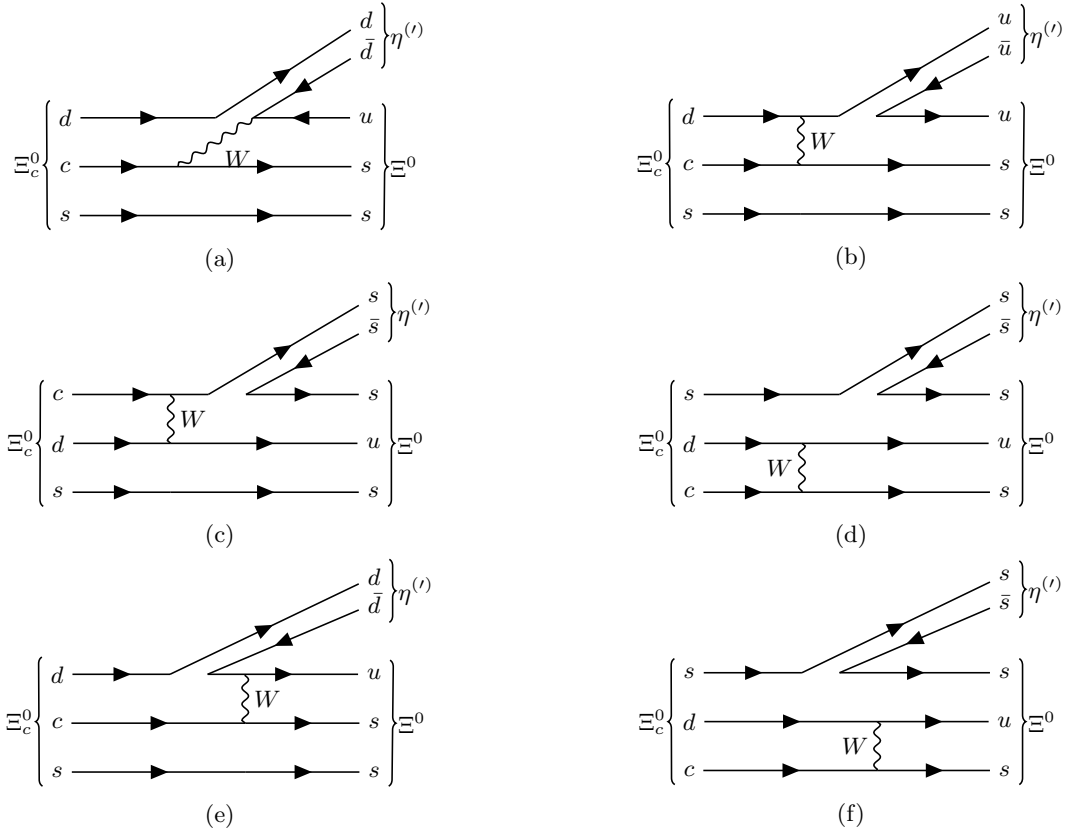

\begin{table}[htbp!]
\centering
\footnotesize
\caption{The amplitudes of $\Xi_c^0\to  \Xi^0 \eta$ (in unit of $10^{-9}\mathrm{GeV}^{-1/2}$ for the real part and $10^{-12}\mathrm{GeV}^{-1/2}$ for the imaginary part). The contributions from the $q\bar q$ for the pole terms, as illustrated by Figs.~\ref{fig:Xizeta}(b) and (e), are listed in the third line and the sixth lines. The contributions from the $s\bar s$ for the pole terms, as illustrated by Figs.~\ref{fig:Xizeta}(c),(d) and (f), are listed in the fourth line and the last lines.}
\begin{ruledtabular} 
\begin{tabular}{clllllllllll}
   &DME &  CS   &\multicolumn{2}{l}{WS}&   \multicolumn{6}{l}{SW}   & Total \\
\hline
   &    &      &$\Xi^0$       &                 &    $\Xi_c^0$ &           &      & $\Xi_c^{'0}$   &    &    &    \\ 
   &\n  & \y   &$2.12-19.10i$ &                 &0             &           &      & $-6.77$        &    &    &   \\
PC &0   &$-4.71$&$2.37-21.33i$ &                 &0             &           &      & $-1.59$        &    &    &$-8.59-40.43i$  \\
\hline
   &      &    &$\Xi^{*0}$    &$\Xi^{0}(1690)$ &  $|^2P_\rho\rangle$  &$|^2P_\lambda\rangle$ & $|^4P_\rho\rangle$ &  $|^2P_\rho\rangle$  & $|^2P_\lambda\rangle$ & $|^4P_\lambda\rangle$ &     \\
   &\n    & \y &$-0.13+1.94i$ &$-0.11+1.10i$   &  $2.39 + 0.51i$         &    0              &  $-3.37-0.70i$     &                        0& $3.37+0.73i$      & $-4.83-1.03i$         &     \\
PV &  0 &$4.62$&$0.63-9.32i$  &$5.07-52.77i$   &  $-0.44 - 0.093i$       &    0              &  $0.66+0.14i$      &                        0& $0.64+0.14i$       & $-0.99-0.21i$         & $7.51-59.57i$  \\
\end{tabular}
\end{ruledtabular}
\label{tab:amp4}%
\end{table}%

\begin{table}[htbp!]
\centering
\footnotesize
\caption{The amplitudes of $\Xi_c^0\to  \Xi^0 \eta'$ (in unit of $10^{-9}\mathrm{GeV}^{-1/2}$ for the real part and $10^{-12}\mathrm{GeV}^{-1/2}$ for the imaginary part). The contributions from the $q\bar q$ for the pole terms, as illustrated by Figs.~\ref{fig:Xizeta}(b) and (e), are listed in the third line and the sixth lines. The contributions from the $s\bar s$ for the pole terms, as illustrated by Figs.~\ref{fig:Xizeta}(c),(d) and (f), are listed in the fourth line and the last lines.}
\begin{ruledtabular} 
\begin{tabular}{clllllllllll}
   &DME &  CS   &\multicolumn{2}{l}{WS}&   \multicolumn{6}{l}{SW}   & Total \\
\hline
   &    &       &$\Xi^0$       &                 &    $\Xi_c^0$ &           &      & $\Xi_c^{'0}$   &    &    &    \\ 
   &\n  &  \y   &$-9.30+83.83i$&                 &0             &           &      & $-1.93$        &    &    &   \\
PC &0   &$-0.52$&$0.52-4.66i$ &                  &0             &           &      & $9.17$        &    &    &$-2.06+79.17i$  \\
\hline
   &    &       &$\Xi^{*0}$    &$\Xi^{0}(1690)$ &  $|^2P_\rho\rangle$  &$|^2P_\lambda\rangle$ & $|^4P_\rho\rangle$ &  $|^2P_\rho\rangle$  & $|^2P_\lambda\rangle$ & $|^4P_\lambda\rangle$ &     \\
   & \n &  \y   &$6.27-93.49i$ &$5.08-52.93i$   &  $0.91 + 0.19i$         &    0              &  $-1.09-0.23i$     &                        0& $1.18+0.26i$       & $-1.42-0.303i$         &     \\
PV &  0 &$0.71$ &$0.47-7.01i$  &$3.81-39.67i$   &  $3.06 + 0.65i$         &    0              &  $-3.91-8.16i$     &                        0& $-4.09-0.89i$      & $5.28+1.12i$         & $16.26 - 193.11i$  \\
\end{tabular}
\end{ruledtabular}
\label{tab:amp5}%
\end{table}%

\begin{figure}[htbp!]
\centering
\begin{subfigure}[htbp! ]{0.45\textwidth}
\begin{tikzpicture}[line width=0.6pt]
\begin{feynman}
\vertex (a1) {$ c $};
\vertex[right=1.5cm of a1] (a2);
\vertex[right=0.5cm of a2] (a3);
\vertex[right=0.5cm of a3] (a4);
\vertex[right=1.5cm of a4] (a5){$ u $};
\vertex[below=2em of a1] (b1) {$ d $};
\vertex[below=2em of a2] (b2) ;
\vertex[below=2em of a5] (b3) {$ u $};
\vertex[below=2em of b1] (c1) {$ s $};
\vertex[below=2em of b2] (c2);
\vertex[below=2em of b3] (c3) {$ s $};
\vertex[above=4em of a5]   (d1) {$s$};
\vertex[above=2.8em of a5] (d2) {$\bar u$};
\diagram* {
(a1) -- [fermion] (a3),
(a4) -- [fermion] (a5),
(b1) -- [fermion] (b2) -- [fermion] (b3),
(c1) -- [fermion] (c2) -- [fermion] (c3),
(a3) -- [fermion] (d1),
(d2) -- [fermion] (a4),
(a2) -- [boson, edge label=$W$] (b2),
};
\draw [decoration={brace}, decorate] (c1.south west) -- (a1.north west)
node [pos=0.5, left] {$\Xi_c^0$};
\draw [decoration={brace}, decorate] (a5.north east) -- (c3.south east)
node [pos=0.5, right] {$\Sigma^+$};
\draw [decoration={brace}, decorate] (d1.north east) -- (d2.south east)
node [pos=0.5, right] {$K^-$};
\end{feynman}
\end{tikzpicture}
\caption{}
\end{subfigure} 
~
\begin{subfigure}[htbp! ]{0.45\textwidth}
\begin{tikzpicture}[line width=0.6pt]
\begin{feynman}
\vertex (a1) {$ s $};
\vertex[right=1.5cm of a1] (a2);
\vertex[right=0.5cm of a2] (a3);
\vertex[right=0.5cm of a3] (a4);
\vertex[right=1.5cm of a4] (a5){$ u $};
\vertex[below=2em of a1] (b1) {$ d $};
\vertex[below=2em of a2] (b2) ;
\vertex[below=2em of a5] (b3) {$ u $};
\vertex[below=2em of b1] (c1) {$ c $};
\vertex[below=2em of b2] (c2);
\vertex[below=2em of b3] (c3) {$ s $};
\vertex[above=4em of a5]   (d1) {$s$};
\vertex[above=2.8em of a5] (d2) {$\bar u$};
\diagram* {
(a1) -- [fermion] (a3),
(a4) -- [fermion] (a5),
(b1) -- [fermion] (b2) -- [fermion] (b3),
(c1) -- [fermion] (c2) -- [fermion] (c3),
(a3) -- [fermion] (d1),
(d2) -- [fermion] (a4),
(c2) -- [boson, edge label=$W$] (b2),
};
\draw [decoration={brace}, decorate] (c1.south west) -- (a1.north west)
node [pos=0.5, left] {$\Xi_c^0$};
\draw [decoration={brace}, decorate] (a5.north east) -- (c3.south east)
node [pos=0.5, right] {$\Sigma^+$};
\draw [decoration={brace}, decorate] (d1.north east) -- (d2.south east)
node [pos=0.5, right] {$K^-$};
\end{feynman}
\end{tikzpicture}
\caption{}
\end{subfigure} 
\\

\begin{subfigure}[htbp! ]{0.45\textwidth}
\begin{tikzpicture}[line width=0.6pt]
\begin{feynman}
\vertex (a1) {$ s $};
\vertex[right=1.5cm of a1] (a2);
\vertex[right=0.5cm of a2] (a3);
\vertex[right=0.5cm of a3] (a4);
\vertex[right=1.5cm of a4] (a5){$ u $};
\vertex[below=2em of a1] (b1) {$ d $};
\vertex[below=2em of a4] (b2) ;
\vertex[below=2em of a5] (b3) {$ u $};
\vertex[below=2em of b1] (c1) {$ c $};
\vertex[below=2em of b2] (c2);
\vertex[below=2em of b3] (c3) {$ s $};
\vertex[above=4em of a5]   (d1) {$s$};
\vertex[above=2.9em of a5] (d2) {$\bar u$};
\vertex[right=2.5cm of a1] (w1);
\vertex[below=2em of w1] (w2);
\diagram* {
(a1) -- [fermion] (a2),
(a3) -- [fermion] (a5),
(b1) -- [fermion] (b2) -- [fermion] (b3),
(c1) -- [fermion] (c2) -- [fermion] (c3),
(a2) -- [fermion] (d1),
(d2) -- [fermion] (a3),
(c2) -- [boson, edge label=$W$] (w2),
};
\draw [decoration={brace}, decorate] (c1.south west) -- (a1.north west)
node [pos=0.5, left] {$\Xi_c^0$};
\draw [decoration={brace}, decorate] (a5.north east) -- (c3.south east)
node [pos=0.5, right] {$\Sigma^+$};
\draw [decoration={brace}, decorate] (d1.north east) -- (d2.south east)
node [pos=0.5, right] {$K^-$};
\end{feynman}
\end{tikzpicture}
\caption{}
\end{subfigure}
\caption{The transition diagrams of $\Xi_c^0\to  \Sigma^+ K^-$. Only the pole terms are allowed and (a) and (b) are also the same at the hadron level.}
\label{fig:SigmaK}
\end{figure}
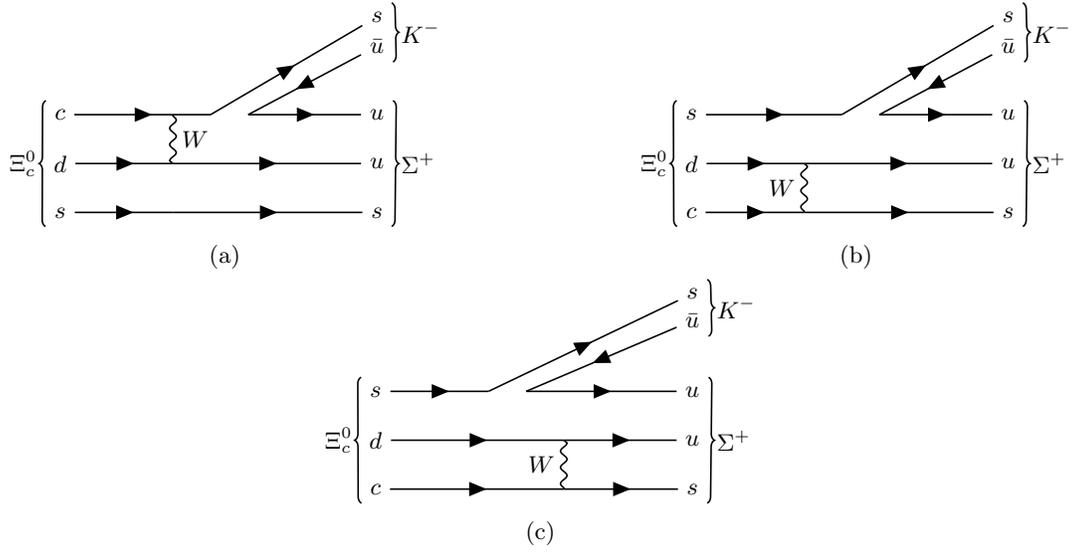

\begin{table}[htbp!]
\centering
\footnotesize
\caption{The amplitudes of the $\Xi_c^0\to  \Sigma^+ K^-$ process (in unit of $10^{-9}\mathrm{GeV}^{-1/2}$ for the real part and $10^{-11}\mathrm{GeV}^{-1/2}$ for the imaginary part).}
\begin{ruledtabular} 
\begin{tabular}{clllllllllll}
   & DME &  CS  & \multicolumn{2}{l}{SW}         &   \multicolumn{6}{l}{SW}   & Total \\
\hline
   & \n  &  \n  & $\Xi^0$        &                 & $\Lambda_c^0$  &           &                 &  $\Sigma_c^{+}$    &   &   &    \\ 
PC &  0  &   0  & $-23.53+21.22i$ &                &  0             &           &                 &  $8.26+2.03i$      &   &   & $-15.27+23.25i$  \\
\hline
   & \n  &  \n  & $\Xi^{*0}$     & $\Xi^{0}(1690)$ &  $|^2P_\rho\rangle$  & $|^2P_\lambda\rangle$ & $|^4P_\rho\rangle$ & $|^2P_\rho\rangle$  & $|^2P_\lambda\rangle$ & $|^4P_\lambda\rangle$ &      \\
PV &  0  &   0  & $1.59 - 2.37i$ & $-1.61 + 1.68i$ &  $3.47 + 0.70i$         &              0     & $-4.88 - 0.96i$    &        0               &  $-4.12 - 0.82i$   & $6.83+1.33i$          & $1.28 - 0.44i$  \\
\end{tabular}
\end{ruledtabular}
\label{tab:amp6}%
\end{table}%

\end{appendix}

\clearpage


\end{document}